\documentclass[preprint,12pt]{elsarticle}

\usepackage{lineno,hyperref}
\modulolinenumbers[1]

\newcommand{\unpara}[1]{\smallskip \noindent \underline {\bf #1:}}

\newcommand\mybox[2][]{\tikz[overlay]\node[fill=blue!20,inner sep=2pt, anchor=text, rectangle, rounded corners=1mm,#1] {#2};\phantom{#2}}

\usepackage{balance}       

\usepackage{graphics}      
\usepackage{graphicx}      
\usepackage[T1]{fontenc}   
\usepackage{savesym}
\usepackage{amsmath}
\savesymbol{iint}
\usepackage{txfonts}
\restoresymbol{TXF}{iint}
\usepackage{quoting}
\usepackage{multirow}
\usepackage{ulem}
\usepackage{color,soul}

\usepackage{multicol}
\usepackage{color}
\usepackage[most]{tcolorbox}
\usepackage{textcomp}
\usepackage{here}
\usepackage{cuted}
\usepackage{capt-of}
\usepackage{microtype}        
\usepackage{ccicons}          
\usepackage{inconsolata}        
\usepackage{booktabs}
\usepackage{subcaption}

\begin{document}

\begin{frontmatter}

\title{Review4Repair: Code Review Aided Automatic Program Repairing}


\author[cdf]{Faria Huq} 
\ead{1505052.fh@ugrad.cse.buet.ac.bd}

\author[cdf]{Masum Hasan} 
\ead{masum@ra.cse.buet.ac.bd}

\author[cdf]{Mahim Anzum Haque Pantho} 
\ead{mahimanzum@gmail.com}

\author[cdf]{Sazan Mahbub} 
\ead{1505020.sm@ugrad.cse.buet.ac.bd}

\author[cdf]{Anindya Iqbal \corref{mycorrespondingauthor}} 
\ead{anindya@cse.buet.ac.bd}
\address[cdf]{Bangladesh University of Engineering and Technology, Bangladesh}

\author[abc]{Toufique Ahmed} 
\ead{tfahmed@ucdavis.edu}
\address[abc]{University of California, Davis, CA, USA}

\cortext[mycorrespondingauthor]{Corresponding author}

\begin{abstract}
\noindent\textbf{Context:} Learning-based automatic program repair techniques are showing promise to provide quality fix suggestions for detected bugs in the source code of the software.
These tools mostly exploit historical data of buggy and fixed code changes and are heavily dependent on bug localizers while applying to a new piece of code. With the increasing popularity of code review, dependency on bug localizers can be reduced. Besides, the code review-based bug localization is more trustworthy since reviewers' expertise and experience are reflected in these suggestions.
\newline
\textbf{Objective:}  The natural language instructions scripted on the review comments are enormous sources of information about the bug's nature and expected solutions. However, none of the learning-based tools has utilized the review comments to fix programming bugs to the best of our knowledge. In this study, we investigate the performance improvement of repair techniques using code review comments.
\newline
\textbf{Method:}  We train a sequence-to-sequence model on 55,060 code reviews and associated code changes. We also introduce new tokenization and preprocessing approaches that help to achieve significant improvement over state-of-the-art learning-based repair techniques. 
\newline
\textbf{Results:} We boost the top-1 accuracy by 20.33\% and top-10 accuracy by 34.82\%. We could provide a suggestion for stylistics and non-code errors unaddressed by prior techniques.
\newline
\textbf{Conclusion:} We believe that the automatic fix suggestions along with code review generated by our approach would help developers address the review comment quickly and correctly and thus save their time and effort.

\end{abstract}


\end{frontmatter}


\section{Introduction}
\label{sec:intro}
\newcommand\tab[1][0.5cm]{\hspace*{#1}}

Code Review has been prevalent in the Software Engineering community for a long time now. In 1976, Fagan introduced a highly structured process for reviewing code \cite{fagan}, which involved intensive line-by-line code inspection. Therefore, it required a significant amount of development time. Over the last few decades, the nature of code review has changed a lot, becoming more informal and tool-based. Big companies such as Microsoft \cite{bosu, bosu-journal}, Google \cite{google-code-review}, Facebook \cite{facebook-code-review} and also the Open Source Software (OSS) projects \cite{developersee, bosu-journal} have adopted lightweight review practice accelerating the review process. Moreover, we are now blessed with some wonderful review tools (e.g., Gerrit, ReviewBoard, Github pull-based reviews, and Phabricator). These tools have enabled the reviewer to make inline comments to enlighten specific issues that the developer needs to address.

\begin{figure}[ht]
\centering
\scalebox{0.9}{
\includegraphics[width=\linewidth]{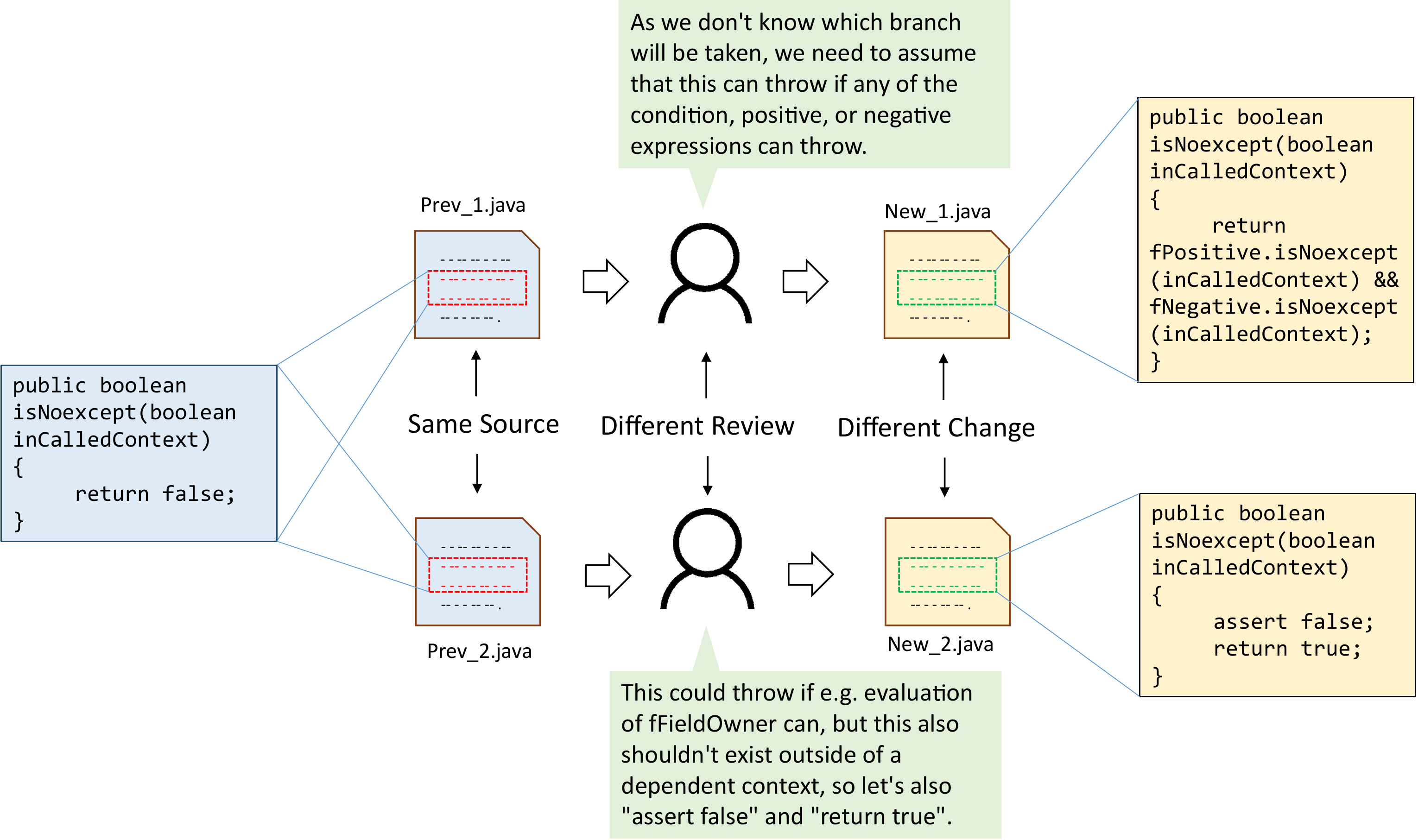}
}
\caption{Impact and importance of code review comment in generating correct code change.}
\label{fig:context}
\end{figure}

Fixing defects in a program are inherently very tedious and expensive, accounting for nearly 50\% of the total cost in Software Development \cite{reversible}. Researchers are trying to provide better solutions by automating these processes \cite{formal-specification, survey}. Classic automatic program repair techniques attempt to modify a program with the help of a specification for the intended program behavior, such as a test suite \cite{semfix, apr1, apr2, apr3, apr4, apr5, apr6, apr7, apr8, apr9}.  Practically, a well-specified test suite is challenging to create, and the generated solutions overfit to a weakly specified test suite~\cite{formal-specification, overfitting1, overfitting2}.  
Recent improvements in advanced machine learning techniques, especially deep learning, and the availability of many patches are encouraging learning-based repair. Instead of relying on a test suite, these techniques rely on previous code fixes in similar code defects. However, they are yet to achieve acceptable quality in most cases.
One beautiful code review attribute is the symbiosis of informal natural comments by the reviewer and a more formal, well-defined structured code authored by the developer.
Can we reduce the developers’ effort by partially automating the bug-fix or refactoring using the review given by the reviewer?
Most of the learning-based repair tools have to depend on bug-localizers to apply fixes to a new code. Code review comments can be utilized as trustworthy bug-localizers because some experience developers verify the bugs' location.  
We envision that code review comments can play an essential role in improving the quality of fix suggestions by providing insight from the reviewer's experience and expertise, localizing the bugs effectively. We illustrate two real-world examples in Figure \ref{fig:context} (taken from Eclipse~\cite{eclipse}) where the source code snippets were similar, but the reviewer's comments were different, which led to two very different solutions.
This paper aims to utilize such communication between the reviewer and the developer, improving state-of-the-art deep learning-based automatic program repair approaches.





\begin{figure}[ht]
\centering
\scalebox{0.9}{
\includegraphics[width=\linewidth]{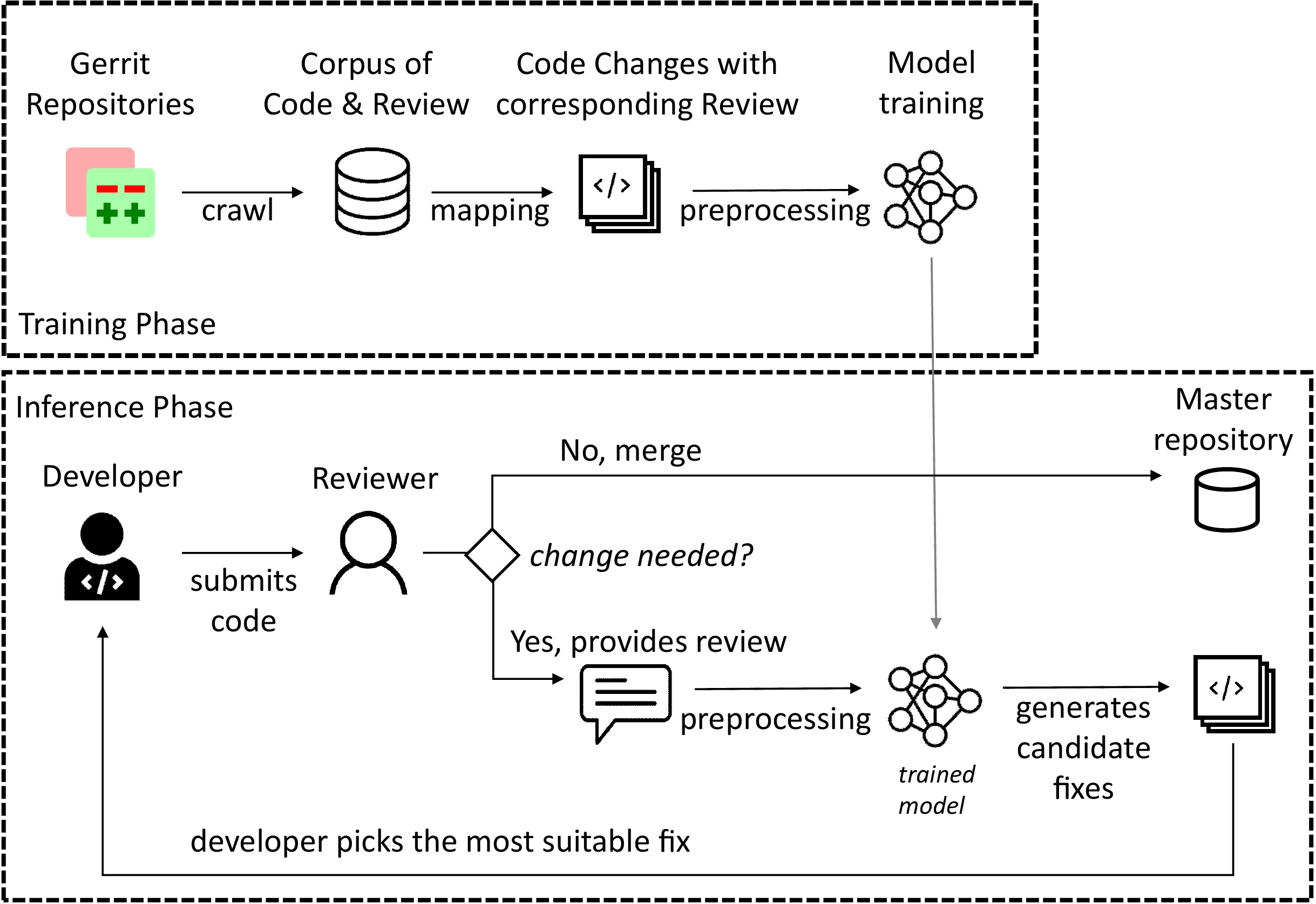}
}
\caption{Complete workflow of our system.}
\label{fig:end_to_end}
\end{figure}

To investigate the impact of code review, We have designed a Neural Machine Translation (NMT) model based on \textit{pointer generator network}~\cite{gehrmann2018bottom} that learns jointly from code review comments written in Natural Language and corresponding code changes. When the reviewer submits a review, the model generates candidate fix suggestions for the intended change. These will be visible to the program author, who can select the best one from the suggestions. Thus, the time and effort needed for the program repair can be reduced, especially for the inexperienced developers who might not be readily aware of the solution.
We conduct several data preprocessing steps, including new tokenization techniques termed as hard and soft tokenization. 
The overall workflow of our system is shown in Figure \ref{fig:end_to_end}.




Our approach covers a wide range of commonly reported issues in code reviews and addresses more types of defects than other works. Specifically, we could suggest the fix for stylistic changes (e.g., indentation and formatting) that increase the readability of the code and non-code issues (e.g., comment, annotation, logs, copyright issues, etc.). 

We also have systematically developed a taxonomy of fixes generated by the tool by studying 501 random samples. 
We have identified four categories (bug fix, refactoring, stylistic change, and non-code change) and 47 sub-categories. 


Our contributions are as follows:
\begin{enumerate}

\item We develop 55,060 training data and 2,961 test data of code changes and their associated code reviews.

\item We develop sequences-to-sequence learning models based on one of the best performing summarization networks followed by extensive preprocessing, new tokenization, and vocabulary creation. We show that utilizing code review and source code improves the repair accuracy by 20.33\% in Top-1 prediction and 34.82\% in Top-10 prediction. Our tool significantly outperforms state-of-the-art learning-based program repair techniques \cite{chen2018sequencer, tufano2019learning}.

\item We provide fixes utilizing code reviews for stylistic and non-code issues along with bug fixes and refactoring, whereas prior works have limited capability to address only the last two types.
We conduct a systematic analysis of 501 randomly selected samples to develop a taxonomy of fixes. We found 47 subcategories of generated fixes depicting our model's ability to learn a wide variety of solutions. 
\end{enumerate}






\section{Motivating Example}
\label{sec:example}
\begin{figure}[ht]
\begin{subfigure}{.5\linewidth}
  \centering
  \includegraphics[width=0.95\linewidth]{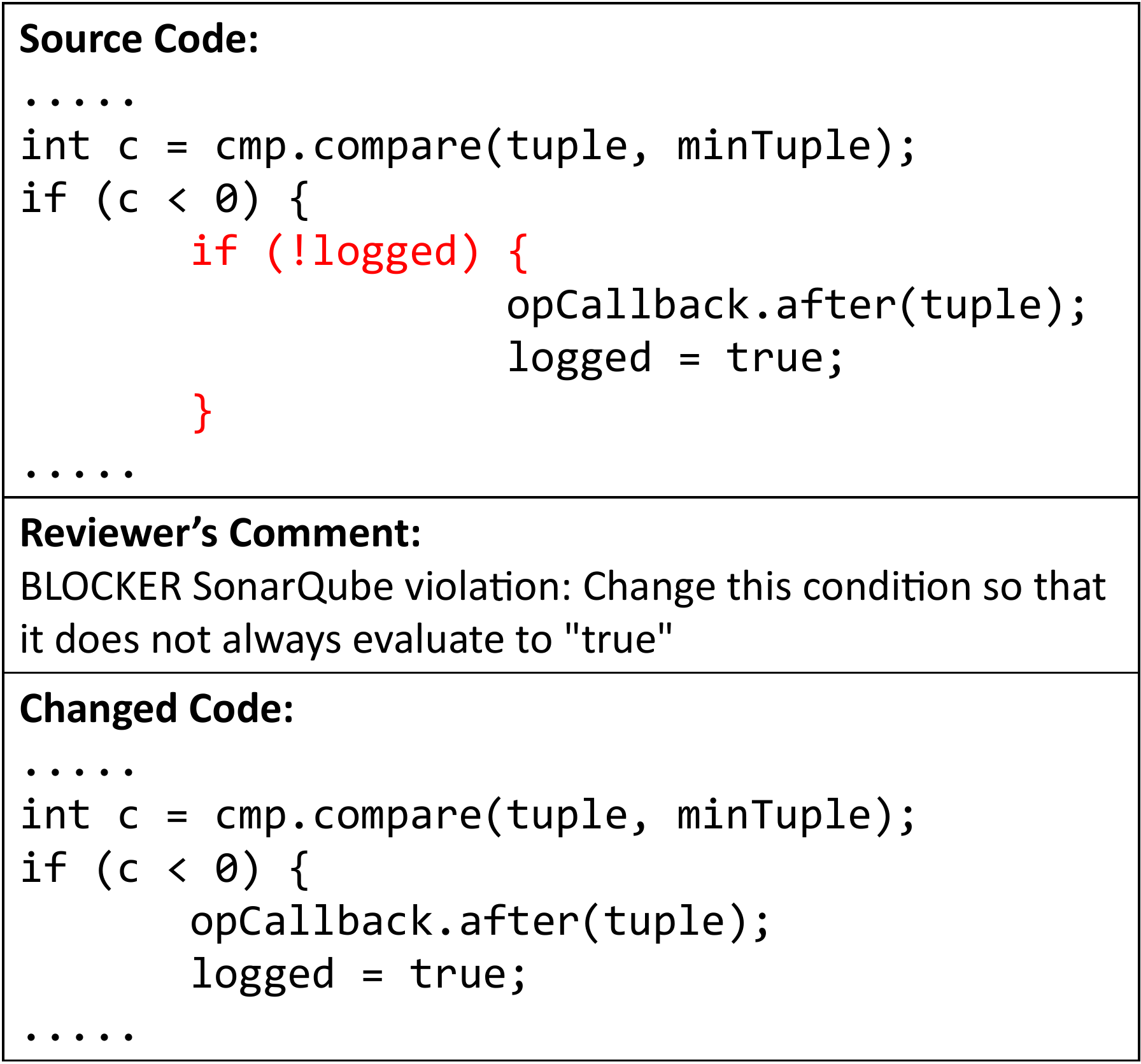}
  \caption{Code before change, review comment from the \\ reviewer and code after change.}
  \label{fig:moti1}
\end{subfigure}%
\begin{subfigure}{.5\linewidth}
  \centering
  \includegraphics[width=0.95\linewidth]{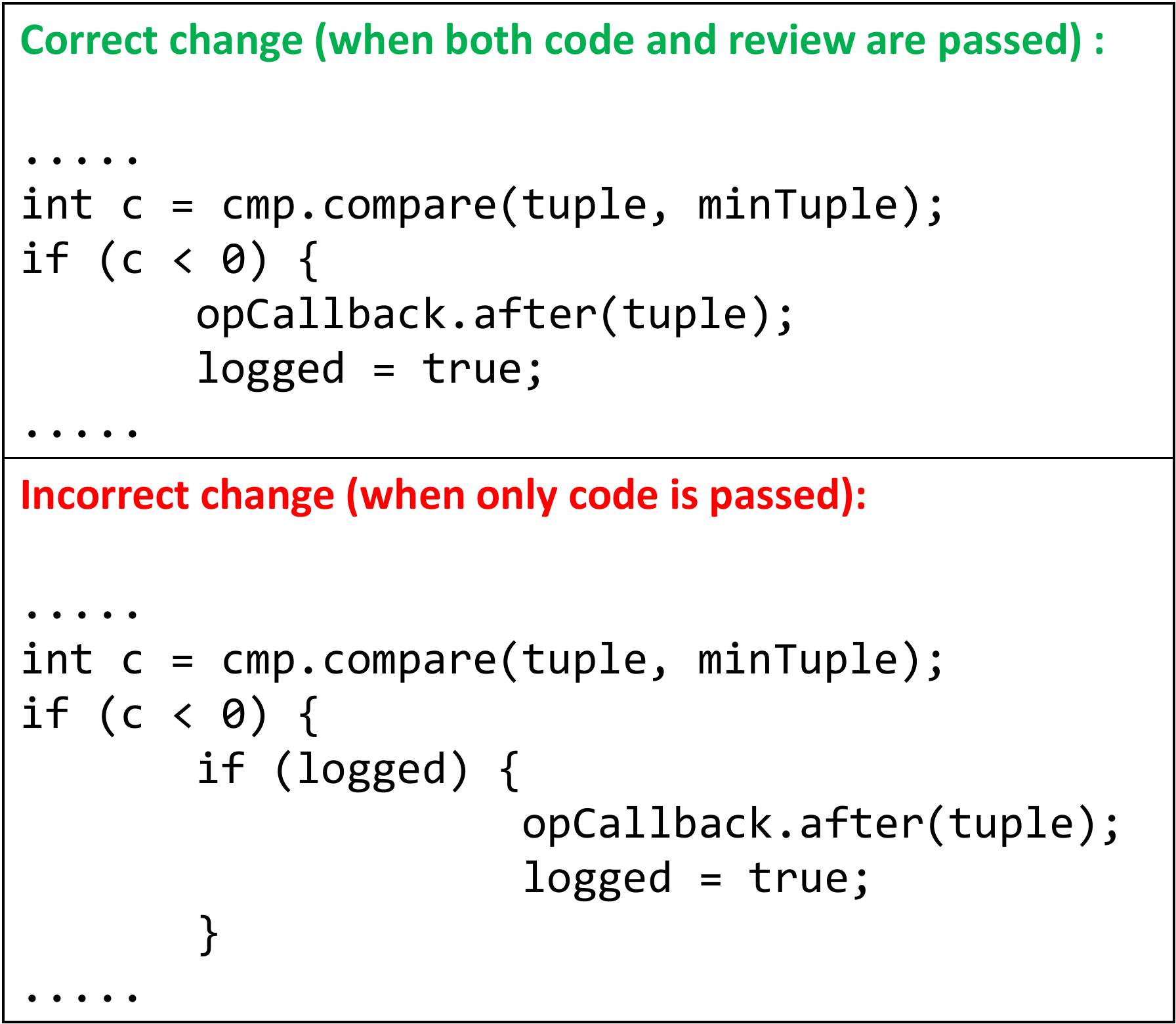}
  \caption{Fix suggestion with and without code review generated by our model. Without code review, correct change was not suggested.}
  \label{fig:moti2}
\end{subfigure}%
\caption{Example of how code review comment may help to generate better fix suggestion.
}
\label{fig:motiv}
\end{figure}

In Figure \ref{fig:motiv}, we illustrate an example demonstrating the utility of our approach. In this scenario, the reviewer commented that the \texttt{if} condition should be changed to prevent always evaluating to the true situation. Given this comment, the developer is supposed to omit the \texttt{if} condition and leave other portions unchanged.
We aim to mimic this activity in our approach, i.e., given the source code and reviewer's comment, our approach should generate the fix that addresses the issues mentioned in the comment. 


This research explores if the quality of fix suggestions can be improved, utilizing the suggestion given in code review comments. First, we built a sequence-to-sequence (seq2seq) network~\cite{gehrmann2018bottom} (see \ref{subsec:network}) using only the code changes and the model achieves comparable performance with state-of-the-art techniques \cite{chen2018sequencer}. We analyzed the examples that could not be properly fixed and considered that there may be some improvements in some cases if the review is also fed to our model. Accordingly, we designed another model that takes review comments as input in addition to code, and we observed improvements in some cases. 
Figure~\ref{fig:motiv} shows one example of an exact fix generated when we pass both the source code and the reviewer's comment. Note that although the generated result without the addition of review comment is correct in logic and syntax, it is not the intended correct solution for the issue mentioned in this particular scenario.
The review comment thoroughly guides this change and any model trained without the review comments is not likely to be able to address this issue.
We show some more examples successfully generated by our model in Table \ref{tab:example_2}.


\begin{table}[ht]
\vspace{-2mm}
\centering
\resizebox{\textwidth}{!}{%
\begin{tabular}{ll}
\hline
\multicolumn{2}{l}{\begin{tabular}[c]{@{}l@{}}\textbf{Code Review:} this will never be thrown on Android (i should go modify the docs to stop claiming this happens)\end{tabular}} \\ \hline
\multicolumn{1}{l|}{\begin{tabular}[c]{@{}l@{}}\textbf{Code Before Change:} \\ 
\texttt{if (oldThread != null) \{} \\ 
~~\texttt{\textcolor{red}{try \{}} \\ 
~~~~\texttt{oldThread.interrupt();} \\ 
~~\texttt{\textcolor{red}{\} catch (SecurityException e) \{}} \\ 
~~~~\texttt{\textcolor{red}{Log.e(TAG, "Interrupting thread", e);}} \\ 
~~\texttt{\textcolor{red}{\}}} \\ 
\texttt{\}}  
\end{tabular}} &
\begin{tabular}[c]{@{}l@{}}
\textbf{Code After Change:} \\ 
\texttt{if (oldThread != null) \{} \\ 
~~\texttt{oldThread.interrupt();} \\ 
\texttt{\}}  
\end{tabular} \\ \hline
\vspace{-4mm}\\ \hline
\multicolumn{2}{l}{\begin{tabular}[c]{@{}l@{}}\textbf{Code Review:} there's no easy way to have Jackson2 omit members based on whatever criteria; we need to understand  \\ a null matrix member as identity.
\end{tabular}} \\ \hline
\multicolumn{1}{l|}{\begin{tabular}[c]{@{}l@{}}\textbf{Code Before Change:}  \\ 
\texttt{public void setMatrix(double[] data) \{} \\ 
~~\texttt{\textcolor{red}{if(data != null) \{}} \\ 
~~~~\texttt{mMatrix = MatrixHelper.createMatrix(data);} \\
~~\texttt{\textcolor{red}{\}}} \\ 
~~\texttt{\textcolor{red}{else \{}} \\ 
~~~~\texttt{\textcolor{red}{mMatrix = MatrixHelper.createMatrix();}} \\ 
~~\texttt{\textcolor{red}{\}}} \\ 
~~\texttt{...\}}
\end{tabular}} & 
\begin{tabular}[c]{@{}l@{}}\textbf{Code After Change:} \\ 
\texttt{public void setMatrix(double[] data) \{} \\ 
~~\texttt{mMatrix = MatrixHelper.createMatrix(data);} \\
~~\texttt{...\}}
\end{tabular}  
\\ \hline
\vspace{-4mm}\\ \hline
\multicolumn{2}{l}{\begin{tabular}[c]{@{}l@{}}\textbf{Code Review:} should this flagged as @Nullable?
\end{tabular}} \\ \hline
\multicolumn{1}{l|}{\begin{tabular}[c]{@{}l@{}}\textbf{Code Before Change:}  \\ 
\texttt{public void addChange(} \\ 
      ~~\texttt{String id, ...,} \\ 
      ~~\texttt{ProjectState projectState, ...)} \\ 
      ~~\texttt{\{...\}}
\end{tabular}} & 
\begin{tabular}[c]{@{}l@{}}\textbf{Code After Change:} \\ 
\texttt{public void addChange(} \\ 
      ~~\texttt{String id, ...,} \\ 
      ~~\texttt{\textcolor{green}{@Nullable} ProjectState projectState, ...)} \\ 
      ~~\texttt{\{...\}}
\end{tabular} \\ \hline
\vspace{-4mm}
\\ \hline
\multicolumn{2}{l}{\begin{tabular}[c]{@{}l@{}}\textbf{Code Review:} let's not modify this. looks like the classes that use this method implement buffered writing
\end{tabular}} \\ \hline
\multicolumn{1}{l|}{\begin{tabular}[c]{@{}l@{}}\textbf{Code Before Change:}  \\ 
\texttt{public OutputStream write(...) throws IOException \{} \\ 
		~~\texttt{File file = getFile(item);} \\ 
		~~\texttt{return \textcolor{red}{new BufferedOutputStream(}new FileOutputStream(file)\textcolor{red}{)};} \\ 
	~\texttt{\}}
\end{tabular}} & 
\begin{tabular}[c]{@{}l@{}}\textbf{Code After Change:} \\ 
\texttt{public OutputStream write(...) throws IOException \{} \\ 
		~~\texttt{File file = getFile(item);} \\ 
		~~\texttt{return new FileOutputStream(file);} \\ 
	~\texttt{\}}
\end{tabular}\\ \hline
\end{tabular}%
}
\caption{Some examples successfully generated by our model when we pass both source code and review comment to it. Deleted and inserted tokens are marked 
by red and green color respectively.}
\label{tab:example_2}
\vspace{-6mm}
\end{table}

\section{Overview}
\label{overview}
In this section, we present an overview of our application scenario and the problem formulation.
\subsection{System Overview}
\label{subsec:overview}

The objective of this study is two-fold. 
\begin{enumerate}
\item Suggesting fixes for the broader range of changes (defects) raised in peer code reviews.
\item Exploiting code review comments written in natural language as an oracle to improve the fix suggestions' quality.
\end{enumerate}

\textcolor{red}{
}

In a code review platform, such as Gerrit, when a developer submits a code patch, a reviewer is assigned and notified to review the code. The reviewer inspects the code, and if the reviewer identifies a defect, he or she highlights one or multiple lines in the code and submits a code review. By \textit{defect}, in this paper, we imply any issue discussed in the review comment that can be related to program functionality, a naming convention, coding style, or even spelling mistakes. The developer addresses the comment and submits a follow-up patch. Finally, when there are no more issues in the code, the reviewer approves the code, and it is merged with the main codebase.

\begin{figure}[ht]
\centering
\includegraphics[width=1.0\linewidth]{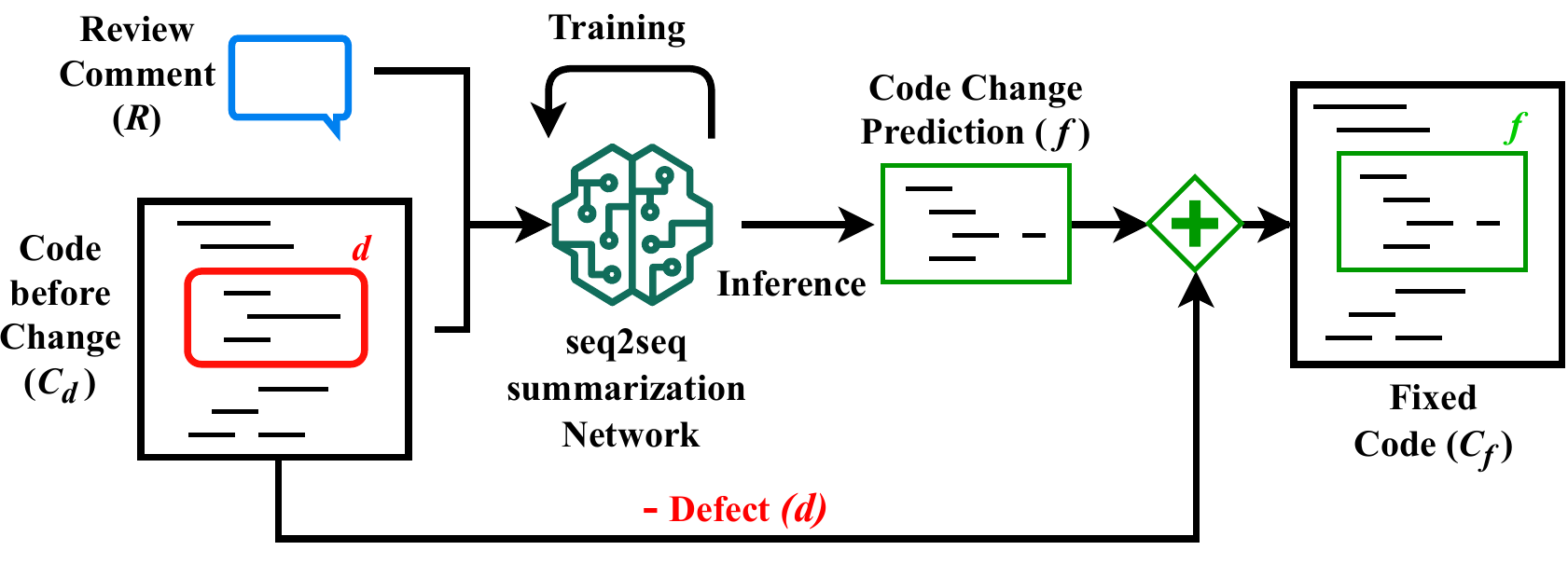}
\caption{Steps showing training of Automatic Program Repair with code review. Model learns to predict code change ($f$), from code before change ($C_d$), defect location, and code review comment ($R$). Replacing the defect $d$ (red round box) with the code change ($f$) creates the fixed code ($C_f$).}
\label{fig:aprpipeline}
\vspace{-4mm} 
\end{figure}

By analyzing these code patches, we can identify the code fragment that was changed due to the code review. In this study, our goal is to create a learning-based system that can predict the changed code automatically by observing a large number of code changes and code review comments in historical data. Once deployed in a production environment, when a reviewer highlights a defect in a code file and writes a review comment, our model will produce multiple fix suggestions for the defect. The developer can choose one of the model's suggestions or write his/her code fix. Figure \ref{fig:aprpipeline} shows different steps of training phase of our model.

\subsection{Problem Definition}
\label{subsec:prob-definition}

We design the task of program repair as a sequence-to-sequence problem. We create two sequence learning models that attempt to repair a defect. Our first model $model\_cc$ is given the code before change $C_d$ and defect location $l$, along with the code review $R$ as input, and tasked to predict the code change $f$ for the defect $d$ (Figure \ref{fig:aprpipeline}).

\unpara{Model I}

\noindent In training time the location $l$ is identified with our `change localization' method (Section \ref{subsubsec:change-localization}). After deployment, we assume that the review comment will localize the bugs, and our deep learning model will fix the error. 

Hence, the prediction by $model\_cc$ can be defined as,
\vspace{2mm} 

\centerline{
$\hat{f}$ = $\underset{f}{\arg \max } \;P\left(f \;|\; C_d, \;l, \;R \right) $
}

\unpara{Model II}

\noindent Our second model $model\_c$ is given the code before change $C_d$, and the location $l$ for defect $d$ as input and tasked to predict the code change $f$. Hence, the prediction by $model\_c$ can be defined as,
\vspace{2mm} 

\centerline{

$\hat{f}$ = $\underset{f}{\arg \max } \;P\left(f \;|\; C_d, \;l \right) $
}

By replacing the defect $d$ from defected code $C_d$ with the fix suggestion $f$, we get the fixed code $C_{f}$.
\vspace{2mm} 

\centerline{$C_{f} $ = $ C_d - d $ + $ f$}
\vspace{2mm}

Using beam search decoding~\cite{rush2013optimal}, the developer will be offered the top $N$ fixed code suggestion $\{C_{f_1}, ... , C_{f_N}\}$ to choose, where $N \in \mathbb N$.

\section{Data Preparation}
\label{sec:data-preparation}
In this section, we briefly describe data collection, data preprocessing, training and test set preparation for this study.

\subsection{Data Collection}
\label{subsec:data-collection}


A learning-based automated code repair approach based on code review requires a large pool of review comments and the associated source code before and after the fix in the training dataset. We choose Gerrit \cite{gerrit} for collecting the data as it is a standard and widely used code review tool.
We created a \textit{GerritMiner} in Java using Gerrit REST API \cite{gerrit-rest-api} and mined 15 Open Source Gerrit repositories consisting of a large number of Java files (Table \ref{tab:project-distribution}). We mined code review comments and associated code files submitted roughly from December 2008 to November 2019. The mining process took approximately 2.5 months on a Intel\textregistered \space Core\textsuperscript{\tiny TM} i7-7700 Processor.

{\small
\begin{table}[ht]
\centering
\begin{tabular}{lrrrr}
\textbf{Project Name} & \textbf{\#CR} & \textbf{\#Java CR} & \textbf{Train} & \textbf{Test} \\ \hline
Acumos~\cite{acumos}	& 6773	& 1387	& 881	& 47	\\
Android~\cite{android}	& 246253	& 23683	& 12512	& 689	\\
Asterix~\cite{asterix}	& 68033	& 23058	& 8509	& 453	\\
Cloudera~\cite{cloudera}	& 151010	& 8623	& 3538	& 197	\\
Couchbase~\cite{couchbase}	& 68864	& 1347	& 808	& 45	\\
Eclipse~\cite{eclipse}	& 51919	& 16903	& 11612	& 621	\\
Fd IO~\cite{fdio}	& 26281	& 866	& 612	& 34	\\
Gerrithub~\cite{gerrithub}	& 116464	& 2102	& 1334	& 66	\\
Googlereview~\cite{googlesource}	& 141410	& 23857	& 13849	& 734	\\
Iotivity~\cite{iotivity}	& 61462	& 1286	& 847	& 48	\\
Others~\cite{omnirom,opencord,polarsys,unicorn,carbonrom} & 10201	& 878	& 558	& 27	\\ \hline
\textbf{Total}	& \textbf{948670} &	\textbf{103990} &	\textbf{55060} &	\textbf{2961}
\end{tabular}
\caption{Project-wise data distribution.}
\label{tab:project-distribution}
\vspace{-6mm} 
\end{table}
}


We mine 1,068,536 code reviews altogether. To ensure that our model is learning only meaningful changes, we carefully discard all code reviews that did not trigger any change near the review comment. We also discard all follow-up conversations to a previous review because they contain incomplete information in our context. Following previous studies in the literature \cite{tufano2019learning, tufano2019empirical, chen2018sequencer, codit, encore}, we intend to work on program repair in Java files only. Hence, for our experiments, we only work on the \textit{.java} files.


\subsection{Noise Removal}
\label{subsec:Noise}
Previous studies~\cite{bosu,developersee,pred_usefulness} show that every code review comment may not always be useful or relevant to the changes. After manual investigation on our dataset, we curate a list of such comments (shown in Table~\ref{tab:irrelevant}) and discard them. 1.32\% inline comments were discarded after this step.


\begin{table}[ht]
\centering
\resizebox{\textwidth}{!}{%
\begin{tabular}{|c|}
\hline
Irrelevant Review Comment                                                       \\ \hline
\begin{tabular}[c]{@{}l@{}}same as above, same as the above, same here, see comment above, same question here,\\ perhaps this as well, see comment above, as discussed, new comment as above, same,\\ see above, similar to above, same concern as above, same comment as above, and here,\\ here too, same comments as above, same thing, same complaint here, same as below,\\ nit, ditto, thanks, fixed with the next upload, uh no, nice, nice thanks, love it,\\ `ok, fixed with next update', `yes,~you are right', done, likewise, i see, and again\end{tabular} \\ \hline
\end{tabular}%
}
\caption{List of Irrelevant Review Comment.}
\label{tab:irrelevant}
\end{table}

\subsection{Input Representation}
\label{subsec:input-representation}
In this section, we discuss how raw source code files were formatted for the learning model. 

\subsubsection{Change Localization:}
\label{subsubsec:change-localization}
\begin{figure}[ht]
\centering
\includegraphics[width=0.8\linewidth]{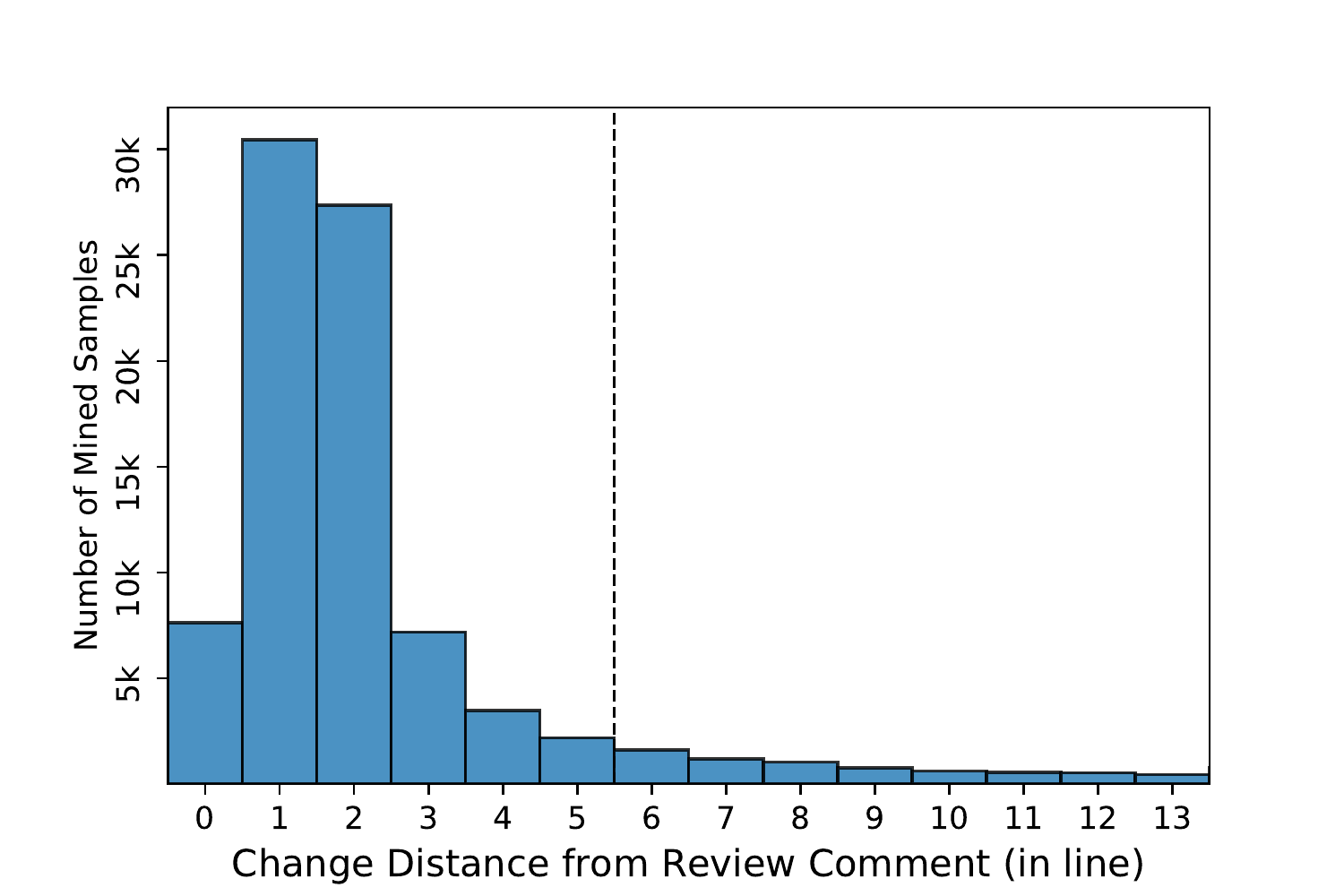}
\caption{Line distance distribution between corresponding line of code review and nearest code change.}
\label{fig:line-cover}
\end{figure}

To identify the exact location of changes in the codes of training data, we build a \textit{``Change Calculation''} tool using \textit{Java DiffUtils} \cite{javadiffutils}. The tool takes the code file before and after the change and calculates the two files' differences. We consider that the code change that is closest to the code review location is a result of the code review. Bosu \textit{et al.} \cite{bosu} demonstrated that useful code review comments trigger a change close to the line where the comment was submitted. We refer this line as \textit{review\_line}. To investigate the case with our dataset, we observe the line difference between \textit{review\_line} and the place of the nearest code change. This is shown in Figure \ref{fig:line-cover}. The distribution shows that 91.27\% of the nearest changes are within 5 line difference from the review\_line. Hence, we consider each change starting within the window of 5 lines to develop training data and discard the sample corresponding to the changes starting outside this window. 



\begin{figure}[ht]
\centering
\includegraphics[width=0.9\linewidth]{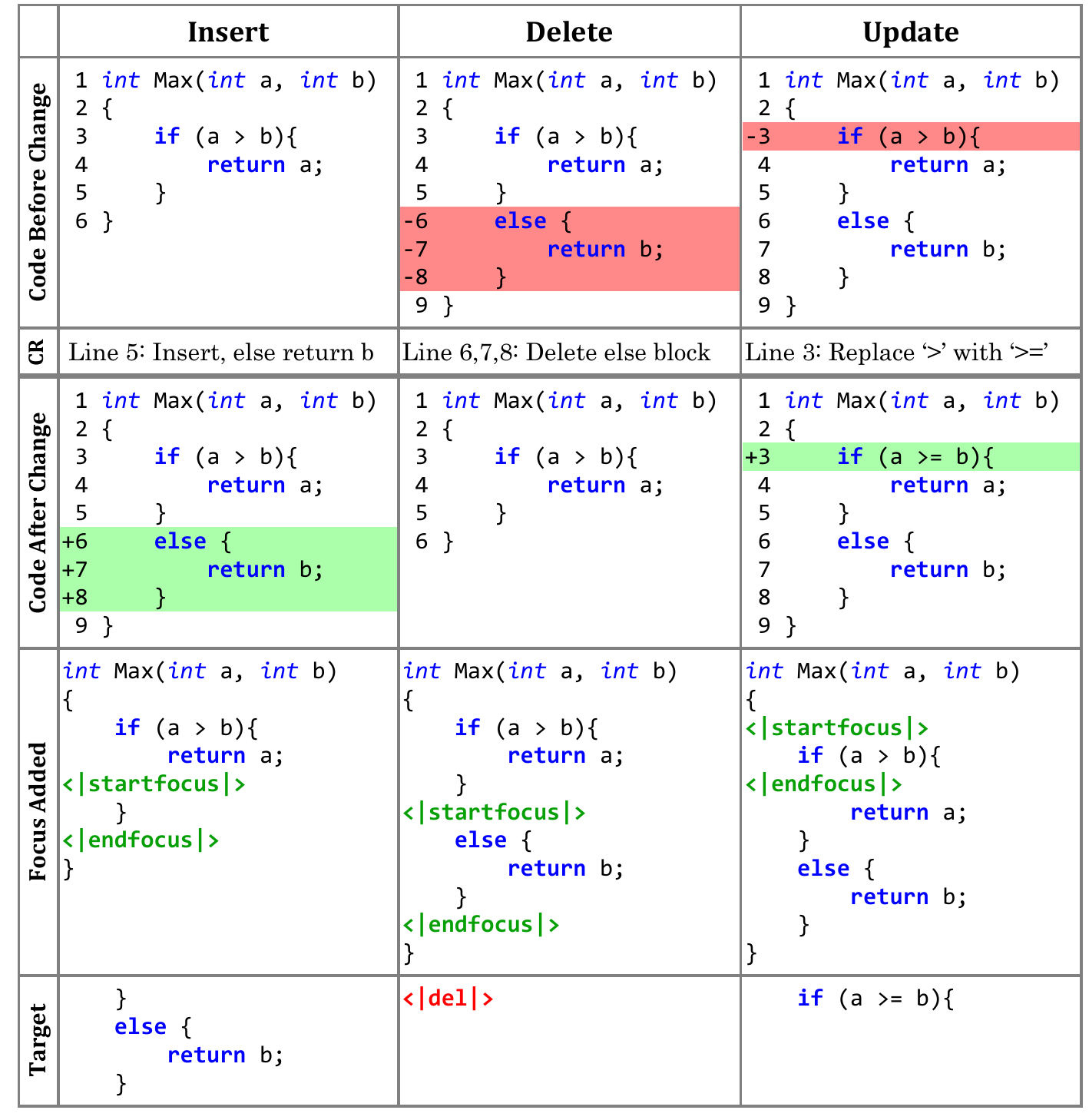}
\caption{Change localization for insert, delete, and update operation.}
\label{fig:change-localization}
\end{figure}

After deciding on the relevancy of code change and review comments, we explicitly concentrate on the code change.
We term the buggy source code as \textit{code before change}, and the lines changed there as \textit{focus}. We mark focus with two special tokens, i.e., \texttt{<|startfocus|>} and \texttt{<|endfocus|>}. 
We call the fixed code as \textit{code after change} and changed portion within the focus as the \textit{target}. 
We elaborate on the terms in an example presented in Figure \ref{fig:change-localization}. Observing the data, the model learns to change the content of the \textit{focus} into the \textit{target} using the code review comment and the surrounding code as context.



We take different measures for identifying the \textit{focus} and the \textit{target} depending on whether the code change is an insert, delete, or update operation. Figure \ref{fig:change-localization} demonstrates these three measures. We design our system so that in a production environment, a reviewer can select one or multiple lines of the code and submit a comment. The model will consider the selected lines as \textit{focus} and try to predict some solutions for it. Replacing the \textit{focus} with one of the predicted solutions will generate a syntactically, semantically, and stylistically correct code. The author will select a suitable one from the predicted solutions. 
Now, we discuss how we deal with the three types of edits.



\begin{enumerate}
    \item \textbf{Insert:} Intuitively, we expect the reviewer to type the comment just before the line where he or she would like the code to be inserted. Hence, we consider the previous line of the insertion operation as the focus. In Figure \ref{fig:change-localization} the reviewer submits a comment on \texttt{Line 5}, suggesting the author to add an else block. Accordingly, the author inserts an else block after line 5. In this case, we consider  \texttt{line 5} as the \textit{focus} and the inserted code, along with the selected line, is considered the \textit{target}.
    
    \item \textbf{Delete:} In delete operation, the code inside \textit{focus} is no longer present in the changed commit. Hence, to indicate deletion, our model produce a special token \texttt{<|del|>} as the \textit{target}. In Figure \ref{fig:change-localization}, the reviewer selects \texttt{Line 6,7,8} and requests the author to delete the else block. Here \texttt{Line 6,7,8} is considered as the \textit{focus} and a special token \texttt{<|del|>} is considered as the \textit{target}.
    
    \item \textbf{Update:} The update operation is more straightforward, i.e., the lines in the original code that require change are considered \textit{focus}, and the corresponding changed lines are considered the \textit{target}.
    
\end{enumerate}

\vspace{-4mm} 


\subsubsection{Code Review Aware Tokenization:}
\label{subsubsec:tokenization}

The commonly used tokenization method (mentioned as \textit{soft tokenization} in this paper) applied in \cite{chen2018sequencer, tufano2019learning, tufano2019empirical, codit} does not contain whitespace tokenization or identifier splitting.  In a shared programming environment, maintaining a consistent style is an essential task for the programmers. Therefore, we propose a tokenization method (named as \textit{hard tokenization}). \textit{Hard tokenization} method has two unique features, as discussed below.

\begin{figure}[ht]
\vspace{-8mm} 
\centering
\includegraphics[width=0.9\linewidth]{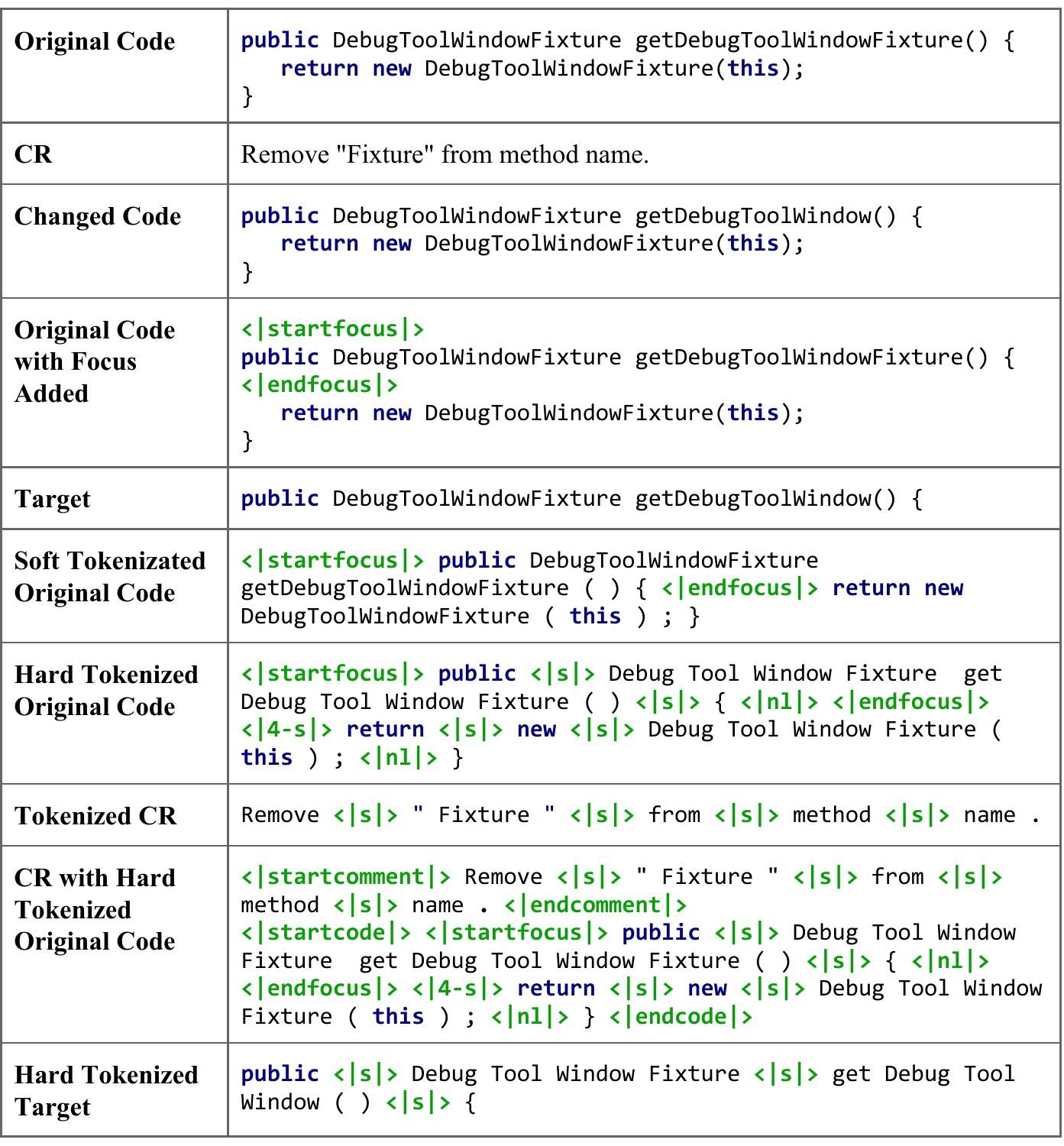}
\vspace{-2mm} 
\caption{Demonstration of \textit{hard} and \textit{soft tokenization} method. \textit{Hard tokenization} splits tokens in atomic Natural Language units, and considers whitespace groups as special tokens.}
\label{fig:tokenization-table}
\vspace{-4mm} 
\end{figure}

\begin{enumerate}

\item \textbf{Whitespace Tokenization:} Programmers frequently use consecutive whitespaces (tabs and space) to indent their code. As we want to preserve these coding styles, we need to consider the whitespaces in our model. However, considering each whitespace character as an individual token would significantly affect the input stream and influence the model's learning process. Therefore, we replace consecutive whitespaces with single tokens.




\item \textbf{Splitting \texttt{camelCase} and \texttt{snake\_case} Identifiers:} Identifier names such as variable, function, or class names contain human language components that carry meaning about the identifier's functionality. Reviewers often make comments about the atomic components, instead of the full identifier name. Hence, we split all \texttt{camelCase} and \texttt{snake\_case} identifiers so our model can identify those atomic components from code and execute the instruction given in the review comment. An example of the splitting process is presented in Figure \ref{fig:tokenization-table}. It also reduces the size of vocabulary~\cite{Allamanis, vocalmodel}. Our dataset's total number of unique tokens was reduced from 1,99,361 to 43,753 (78.05\% reduction) due to identifier splitting.


\end{enumerate}

We implement it using multiple tokenizers used in NLP ( TweetTokenizer, WordPunctTokenizer, MWETokenizer in NLTK Library \cite{nltk}), which allow us to tokenize both code and code review in the same format.


\vspace{-2mm} 

\subsubsection{Input Sequence:}
\label{subsubsec:input-sequence}
In our proposed design, an essential task would be to provide the code change's surrounding context to the learning model. The ability to provide the right context would help the model understand the defect better, copy tokens from the surrounding code, reduce overfitting, and improve generalization. It also needs to consider the following goals:
\begin{enumerate}
\item Reduce code and code review into a reasonably concise sequence of tokens as a sequence to sequence neural network suffers from a long input size.
\item Subsume as much useful information as possible to allow the model to capture the context better.

\end{enumerate}



We feed a context of window size $W$ to the model. The context consists of the $focus$ and its surrounding tokens. We apply the following rules to generate the context for a review comment.  

\begin{enumerate}
\item If the \textit{focus} is written within a function scope and the function is smaller than $W$ tokens, we consider the entire function as input.
\item If the focus is inside a function scope and the function is larger than W tokens, we keep up to W tokens (up to W/2 tokens from the preceding part of the focus and W/2 tokens from the focus and subsequent part) within that function scope as input. 
\item If the \textit{focus} in the global scope, we can follow a similar strategy, i.e., taking up to $W/2$ tokens from the preceding part of the focus and $W/2$ tokens from the focus and subsequent part for input.

\end{enumerate}

\begin{figure}[ht]
\begin{subfigure}{.333\linewidth}
  \centering
  \includegraphics[width=\linewidth]{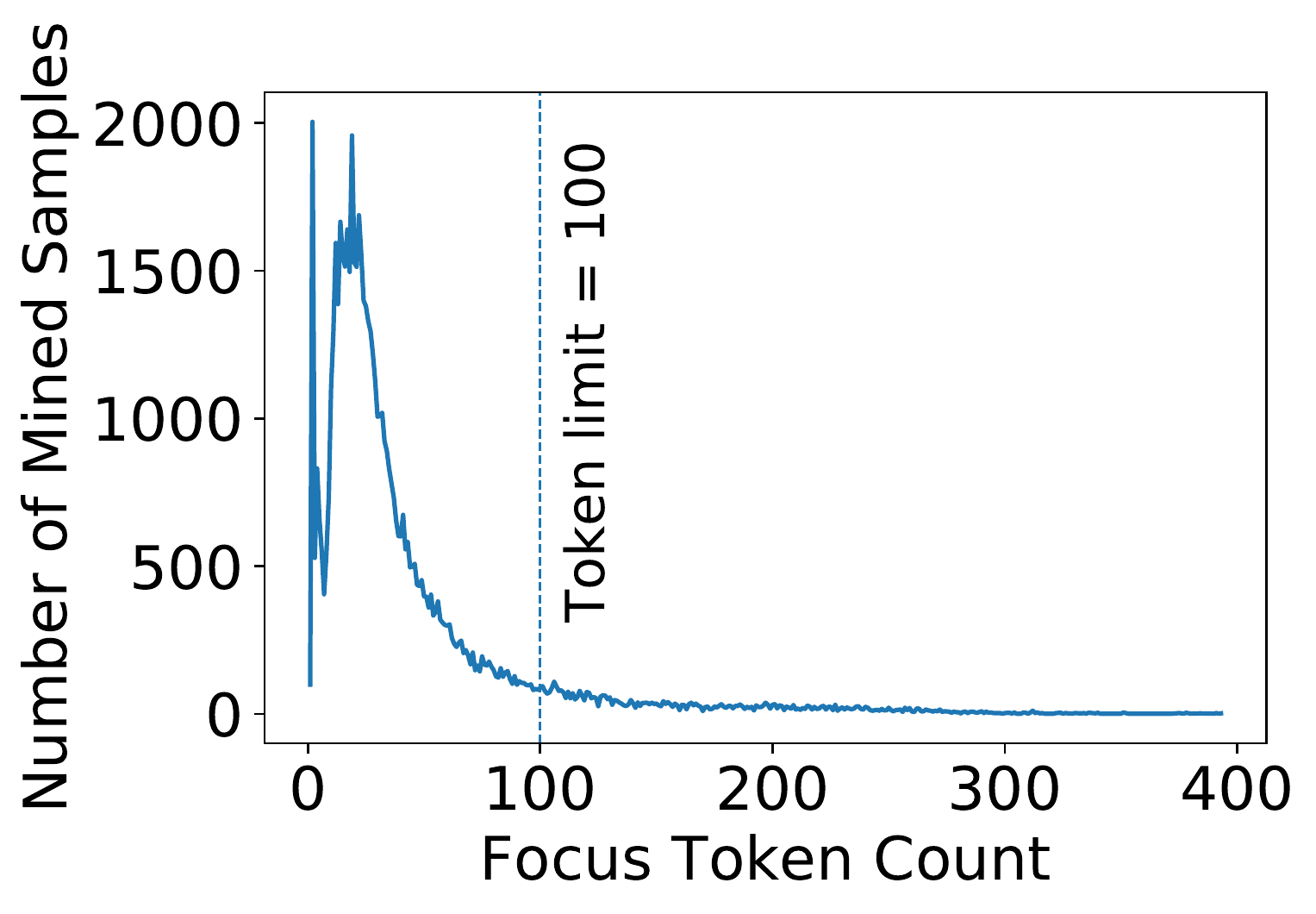}
  \caption{Focus Token Size Distribution}
  \label{fig:input-focus}
\end{subfigure}%
\begin{subfigure}{.333\linewidth}
  \centering
  \includegraphics[width=\linewidth]{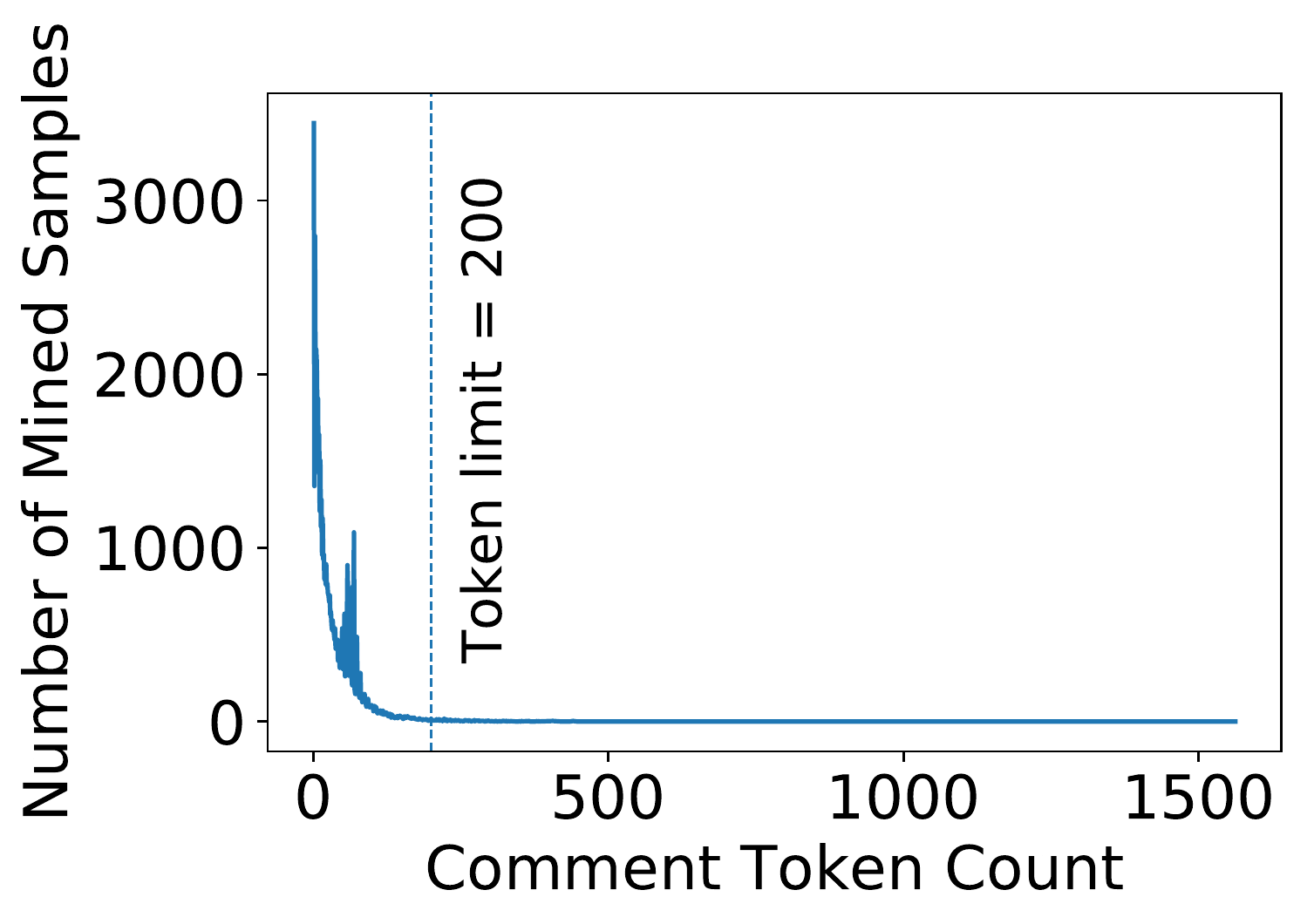}
  \caption{Comment Token Size Distribution}
  \label{fig:comment-len}
\end{subfigure}%
\begin{subfigure}{.333\linewidth}
  \centering
  \includegraphics[width=\linewidth]{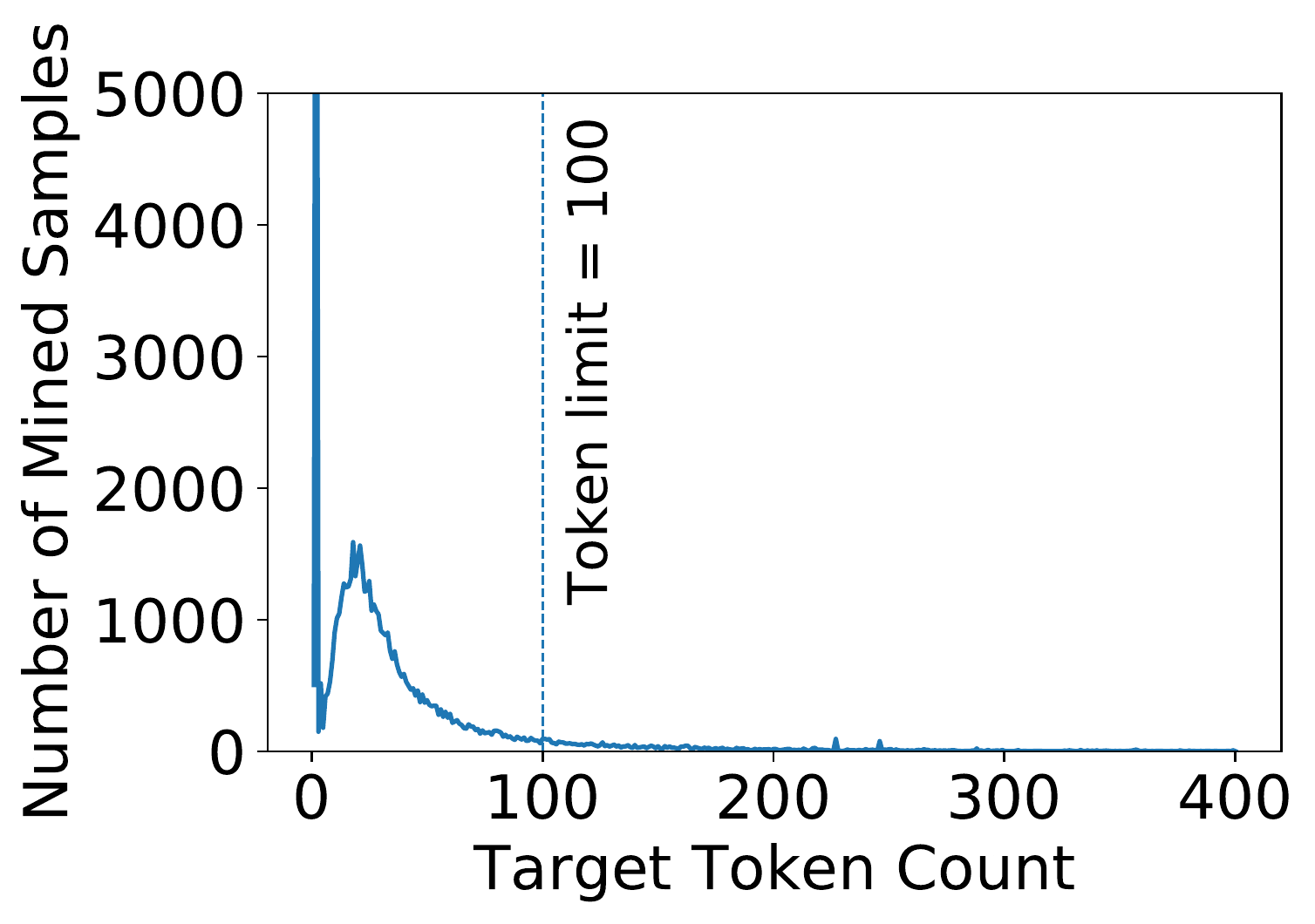}
  \caption{Target Token Size Distribution}
  \label{fig:target-len}
\end{subfigure}%
\caption{Token Size Distribution in our dataset. Our token size limit covers majority of our dataset.}
\label{fig:token_limits}
\end{figure}

The seq2seq network structure we adopt (Section~\ref{subsec:network}) commonly uses 400 to 800 tokens in the input sequence and 100 tokens as output for similar applications such as code summarization \cite{see2017get, paulus2017deep}. We limit the context window $W$ of code to 400. Another element of input sequence, i.e., code review comments are within 200 tokens in 98.725\% cases in our dataset (Figure \ref{fig:comment-len}). Hence, we limit comment size as 200 tokens and consider the first 200 tokens of comments only if it exceeds this length. Thus, the input sequence reaches up to 600 tokens when comments are added to code. We empirically observed that longer sequences result in deteriorating performance. Figure \ref{fig:token_limits} also presents the distribution of focus length and target length.



\subsection{Test Set Generation}
\label{subsec:testdata}

We created a standard test dataset to evaluate different models with different parameters and settings on the same ground. The dataset is suitable to be tested by the models with code review, without code review, and with hard and soft tokenization. We removed all duplicate data points from the dataset and sorted our data in reverse chronological order of the comment submission time. Then we selected the most recent 5\% data from each project. This ensures that our reported performance represents our model's ability to predict future code changes by learning the past code change patterns. Also, by taking 5\% data from each project, we ensure that the test set represents all projects in the dataset. The size of the test data collected from each project is shown in Table \ref{tab:project-distribution}.



\section{Proposed Neural Network Architecture}
\label{sec:neural-network}

In this section, we discuss the Neural Machine Translation (NMT) model that learns code transformation both with and without peer code review. We also describe the NMT model's changes to incorporate both programming language and natural language in the same input stream.

\subsection{Pointer Generator Architecture}
\label{subsec:network}

We train a \textit{pointer generator network} as implemented by Gehrmann \textit{et al.}~\cite{gehrmann2018bottom} available in OpenNMT Pytorch distribution~\cite{opennmt}, which is a state-of-the-art sequence-to-sequence (seq2seq) architecture for text summarization. Seq2seq  networks used for summarization can generate desired output text by deriving and summarizing information from the most relevant parts of the input (i.e., code and review comment). This architecture's ability to generate desired text/code is already established in literature \cite{chen2018sequencer, tufano2019learning, codit}. However, they applied the seq2seq  model only for source code, whereas we have adopted both the source code and natural language code reviews. For this purpose, we incorporate a custom vocabulary as detailed later.

We apply LSTM~\cite{hochreiter1997long} as base RNN with attention mechanism \cite{bahdanau} and copy mechanism \cite{copy} for handling out-of-vocab (OOV) tokens. Programming languages generally result in a large vocabulary because of arbitrary identifier names~\cite{infinitevar}. The basic seq2seq network fails to manage this. In previous studies \cite{chen2018sequencer,codit}, copy mechanism \cite{copy} has been proven effective to overcome the \textit{large vocab problem}~\cite{infinitevar}. When the seq2seq decoder faces an OOV token, the copy mechanism enables the decoder to copy the token from the input token stream directly and place it to the output token stream.

In the original implementation of the pointer generator network, coverage mechanism \cite{coverage} is used to limit the repetition of the tokens in network output. However, programming language keywords repeat frequently. Hence, the use of a coverage mechanism affects program repair (Table \ref{tab:ablation}). Therefore, we exclude it from our recommended model presented in the next subsections.





\subsection{Augmented Vocabulary}
\label{subsubsec:vocab}
Our model uses the separate vocabulary for source and target because it has to encode information from both code and natural language code review comments but generate only code tokens. 
Moreover, code review has a less dominant presence in our input sequence. The average length for a code review and source code in our dataset are 36.80 and 320.38 tokens, respectively. If we apply the standard procedure to create vocabulary followed in the literature~\cite{cho2014learning, cho2014properties, bahdanau, see2017get, dos2014deep, chen2018sequencer, tufano2019learning, tufano2019empirical, bader2019getafix, codit}, most code review tokens will be considered as out-of-vocabulary (OOV) tokens. Hence, our model will fail to gain a contextual understanding of the code review instructions, even with a copy mechanism \cite{copy}. To combat this issue, we propose a larger vocabulary for comments compared to code segments. We consider various combinations of vocabulary size as discussed in Section \ref{sec:ablation_study}. Similar to SequenceR \cite{chen2018sequencer}, we find that large code vocabulary affects model performance. We have experimented with different combinations of vocab sizes (see Section \ref{sec:ablation_study}) and found the best configuration containing the most frequent 2,000 tokens from the codes 
and 8,000 tokens from the comments. 
They cover 93.56\% and 98.86\% tokens out of total source code and code review comment tokens, respectively.

\subsection{Training with Code Change and Code Review Comment (\texorpdfstring{$model\_cc$}{Lg})}
\label{subsec:model-cc}

We create a baseline model with the pointer generator network~\cite{gehrmann2018bottom} that takes both the code (before change) and the code review comment as an input. This model is termed as $model\_cc$. To separate review comment and code from each other, we wrap the review comments with two special tokens \emph{<|startcomment|>} and \emph{<|endcomment|>} and the code with special tokens \emph{<|startcode|>} and \emph{<|endcode|>}. Finally, we concatenate them to produce a single input stream. As discussed earlier in Section~\ref{subsubsec:input-sequence}, the code review and code are limited to 200 tokens and 400 tokens, respectively. Thus the network has an input size of 600. The input vocabulary of baseline $model\_cc$ contains 10,000 tokens; 2,000 are from code and 8,000 are from review comment as described in \ref{subsubsec:vocab}. The output of $model\_cc$ has a maximum length of 100 tokens and 2,000 vocabularies.


\subsection{Training with Code Change Only (\texorpdfstring{$model\_c$}{Lg})}
\label{subsec:model-c}

We create a second baseline model with the pointer generator network~\cite{gehrmann2018bottom} that predicts code change by watching only the code before the change. This network is termed as $model\_c$. Since this model does not consider the review comment, it has to deal with smaller input vocabulary and input sequences than $model\_cc$. Specifically, the network's input code and output are limited to 400 tokens and 100 tokens, respectively. The most frequent 2,000 code tokens in the training dataset are considered for both the input and output vocabulary of $model\_c$. The output of $model\_c$ is identical to $model\_cc$.

\subsection{Inference and Detokenization}
\label{subsec:inference}
After training the model, we use the trained model to generate suggestions. During inference, we prepare our input following hard tokenization, as discussed in Section \ref{subsubsec:input-sequence}. We use beam search decoding~\cite{rush2013optimal} to generate multiple possible suggestions similar to previous works~\cite{tufano2019learning,chen2018sequencer,tufano2019empirical}. We generate our target patches by detokenizing the suggestions from the model. Our \textit{Hard Tokenization} method prevents any information loss. Thus, the source code can be reproduced trivially, preserving whitespace, indentation, and coding style from the token stream.



\section{Experimental Setup}
\label{sec:experiment}


This section describes our neural network model's specific implementation details, evaluation criteria, and comparison method for state-of-the-art models.

\subsection{Evaluation Criteria}
\label{subsec:evaluation}
We evaluate each of our models in the standard test set $T$ described in Section \ref{subsec:testdata}. For each $t \in T$  we perform inference with Beam Search Decoding~\cite{rush2013optimal} with beam size $k$=$10$, which was commonly used in literature \cite{tufano2019empirical} and our empirical finding also supports it. We measure the top-1 accuracy, i.e., the percentage of fixes that our model predicts as the
top-most suggestion, top-5 accuracy, i.e., the percentage of fixes that it predicts as one of the first five suggestions, and similarly measure top-10 accuracy.

We further manually analyzed the predictions made by the models and evaluate their quality for different types of code changes (\ref{tab:taxonomy}).

\subsection{Network Parameters}
\label{subsec:Parameters}
We experiment with our model with different parameter settings and justify the choices in an ablation study (Section \ref{sec:ablation_study}). The best performance is obtained with the following model architecture.


\begin{itemize}
    \item Input Embedding: 2002$\times$256 ($model\_c$), 10002$\times$256 \\  ($model\_cc$); 2,000 and 10,000 vocabulary and 2 special tokens each.
    \item Input sequence length: 400 ($model\_c$), 600 ($model\_cc$)
    \item Output sequence length: 100 (both $model\_c$ and $model\_cc$)
    \item Encoder Bidirectional LSTM size: 256$\times$128$\times$2
    \item Bridge between Encoder and Decoder: 128$\times$128$\times$2 + 128$\times$2
    \item Decoder LSTM size: 512$\times$256
    \item Global Attention: 256$\times$256$\times$3 + 512$\times$256 + 256$\times$2
    \item Token Generator Decoder: 2000$\times$256
    \item Copy Generator: 256$\times$2000 + 2000 + 256$\times$1 + 1
    \item Coverage Attention: \emph{False}
    \item Beam size during inference: 10
\end{itemize}



\subsection{Hardware and Training Time}
We trained our model on NVIDIA\textsuperscript{\tiny\textregistered} V100 Tensor Core GPU with 16 GB VRAM, 16GB RAM and eight-core CPU on Google Cloud Platform. Training each sequence to sequence model up to 80,000 training steps took nearly 72 hours of training time. 

\subsection{Comparison with State-of-the-Art Models}
\label{subsec:sota}
In this section, we discuss the methodology of comparing our models with two recent comparable works~\cite{tufano2019learning, chen2018sequencer}.

Tufano \textit{et al.}~\cite{tufano2019learning} create two different neural machine translation models, one with functions less than 50 tokens, and the other with functions between 50 to 100 tokens in \textit{soft tokenization}.
The dataset for their study is collected from three large code repositories: Android \cite{android}, Google Source \cite{googlesource}, and Ovirt \cite{ovirt}. One of these (Android) is common with our dataset (Table \ref{tab:project-distribution}). Since our model requires review comments, it is infeasible to use the exact dataset proposed by them \cite{tufano2019learning}. Therefore, we decided to use test data extracted from Android project only for a fair comparison. We created two code change test datasets from Android project with two settings of token size, as mentioned below: 


\begin{enumerate}
\item $Test_{small}$: containing 292 instances where token count is: $0 < token\_ count \leq 50$; 
\item $Test_{medium}$: containing 246 instances where token count is: $50 < token\_ count \leq 100$. \end{enumerate}

Both of these test datasets contain only functions with a single code review comment and a single code change. These data points are carefully removed from our training dataset. We reproduce the two models proposed by Tufano \textit{et al.} \cite{tufano2019learning} with the data and source code released by the authors~\cite{tufano-code} and achieve nearly identical validation result as reported in their paper. After validating their model, we compare our approach with them by training their model with our dataset. The results comparing Tufano \textit{et al.} and our models on $Test_{small}$ and $Test_{medium}$ datasets are shown in Table \ref{tab:tufano}.

SequenceR \cite{chen2018sequencer} performs single line update operations inside functions with defects, given the code of the function and the line number of the defect. Their model cannot handle insert, delete, multi-line operations, and defects outside of function scope (i.e., for comments and global data). To make a comparison with SequenceR \cite{chen2018sequencer}, we selected 349 instances from our standard test dataset (Section \ref{subsec:testdata}) that comprise single line update operations only. We implemented SequenceR preprocessing, training, and test pipeline with the help of their source code \cite{sequencer-code} and achieve similar performance with the reported performance of in their paper. We created two different implementations of SequenceR after validating the original work. The first model was trained with the 35,578 training data provided with their paper and termed as $SequenceR$.The second model is trained with training data collected from our mined data that satisfy the SequneceR dataset constraints. We have trained it with 56,000 data to make a fair comparison with our model. This implementation of SequenceR is referred as $SequenceR_{new}$. We test both these models and our best models with and without code review on the prepared test set of 349 data. The comparison is shown in Table \ref{tab:sequencer}.

\section{Evaluation and Results}
\label{sec:results}

In this section, we discuss the experimental results and findings of our research.

\subsection{How effective is code review in automatic code repair?}
\label{subsec:hypothesis}

In this study, we aim to show whether code review can improve automatic code repair performance. We train our model with two different settings, with and without code review termed as \textit{\textbf{$model\_{cc}$}} and \textit{\textbf{$model\_{c}$}}, respectively. The construction of these models is discussed in Section \ref{subsec:network}.


\begin{figure}[ht]
\includegraphics[width=1\linewidth]{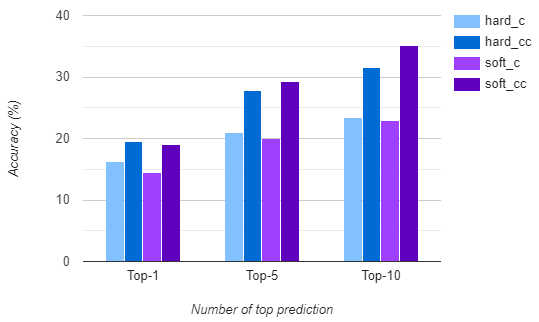}
\caption{Top-1, top-5, and top-10 test accuracy of model trained without code review (c) and with code review (cc) for both hard and soft tokenization method.}
\label{fig:hard-soft-c-cc}
\end{figure}


Figure \ref{fig:hard-soft-c-cc} and Table~\ref{tab:baseline} clearly show that incorporating the code review comments improve the prediction accuracy for both \textit{hard tokenization} and \textit{soft tokenization} methods in all top-1, top-5, and top-10 predictions. Since Hard Tokenization has better top-1 accuracies on both \textit{\textbf{$model\_{cc}$}} and \textit{\textbf{$model\_{c}$}}, we decided to use it for further analysis.






\begin{table}[ht]
\centering
\begin{tabular}{lccc}
\toprule
\textbf{Model} & \textbf{Top-1} & \textbf{Top-5} & \textbf{Top-10} \\ \midrule
\textbf{Baseline \textit{model\_c}}      & 16.29                               & 20.94          & 23.37           \\ 
\textbf{Baseline \textit{model\_cc}}     & 19.59                               & 27.73          & 31.51           \\ 
\textbf{Relative improvement}      & \textbf{+20.33}                               & \textbf{+32.41}          & \textbf{+34.82} \\ 
\bottomrule
\end{tabular}
\caption{Baseline model accuracy (in percent) for \textit{model\_c} and \textit{model\_cc} in \textit{hard tokenization}, and relative improvement of \textit{model\_cc} over \textit{model\_c}.}
\label{tab:baseline}
\end{table}

\subsection{How effectively does the model perform in comparison with state-of-the-art techniques?}
\label{subsec:ablation}

We aim to evaluate our model on a benchmark against well-established approaches \cite{chen2018sequencer, tufano2019learning}. 
To ensure that we replicate the exact settings used by previous architectures, we generate separate test cases for comparing our models, as described in Section \ref{subsec:testdata}.

\subsubsection{Comparing with the methodology proposed by Tufano \textit{et al.} \texorpdfstring{\cite{tufano2019learning}}{Lg}}
As discussed earlier in Section~\ref{subsec:sota}, We show the comparison among Tufano \textit{et al.}~\cite{ tufano2019learning} and our models on $Test_{small}$ and $Test_{medium}$ in Table~\ref{tab:tufano}. The result shows both {$model\_{c}$}  and {$model\_{cc}$}  outperform Tufano et al~\cite{ tufano2019learning} on both test sets.


\begin{table}[ht]
\centering
  \begin{tabular}{cccl}
    \toprule
    Model & Top n Prediction & $Test_{small}$(292) & $Test_{medium}$(246)\\
    \midrule
    \multirow{3}{3em}{Tufano \textit{et al}.\cite{tufano2019learning}} & 1 & 2 (0.68\%) & 1 (0.41\%) \\ 
    & 5 & 6 (2.05\%) & 3 (1.22\%) \\ 
    & 10 & 7 (2.40\%) & 4 (1.63\%) \\ 
    \midrule
    \multirow{3}{3em}{$model\_{c}$} & 1 & 21 (7.19\%) & 11 (4.47\%) \\ 
    & 5 & 52 (17.80\%) & 38 (15.44\%) \\ 
    & 10 & 80 (27.40\%) & 46 (18.69\%) \\ 
    \midrule
    \multirow{3}{3em}{$model\_{cc}$} & 1 & 31 (10.61\%) & 24 (9.76\%) \\ 
    & 5 & 71 (24.31\%) & 55 (22.36\%) \\ 
    & 10 & 93 (31.85\%) & 63 (25.61\%) \\ 
  \bottomrule
  \end{tabular}
  \caption{Comparison of Tufano \textit{et al}.\cite{tufano2019learning} and our models.}
  \label{tab:tufano}
\end{table}

\subsubsection{Comparing with the methodology proposed by Chen et al. \texorpdfstring{\cite{chen2018sequencer}}{Lg}}

\begin{table}[ht]
\centering
\begin{tabular}{lccc}
\toprule
\textbf{Model}                     & \textbf{Top 1} & \textbf{Top 5} & \textbf{Top 10} \\ \midrule
SequenceR$_{original}$ & 1.27\%         & 1.52\%         & 2.02\%          \\ 
SequenceR$_{new}$            & 3.03\%         & 6.58\%         & 7.34\%          \\ 
\textbf{$model\_c$}             & 3.54\%         & 12.91\%        & 16.96\%         \\ 
\textbf{$model\_cc$}            & 8.86\%         & 18.48\%        & 25.31\%         \\ 
\bottomrule
\end{tabular}
\caption{Comparison with SequenceR \cite{chen2018sequencer}.}
\label{tab:sequencer}
\end{table}

We evaluate two different implementations of SequenceR, as mentioned in Section~\ref{subsec:sota}. 
First, we apply the SequenceR model, released by the authors, and evaluate it on the test data (named it as SequenceR$_{original}$). To make a fair comparison, we also create a training data set for SequenceR from our corpus and train the SequenceR again ( named as SequenceR$_{new}$). Both of the results are displayed in Table \ref{tab:sequencer}.

We can see that the original SequenceR model performs very poorly on our test set. This poor performance is attributed to the difference in the vocabulary of the training and test dataset. Our models perform significantly better than SequenceR$_{new}$. We can see that the top-1 prediction of our $model\_c$ is comparable to the performance of SequenceR$_{new}$. However, the top-1 prediction of $model\_cc$ is significantly better because of the addition of code review comments.



\begin{figure}[ht]
\centering
\includegraphics[width=0.7\linewidth]{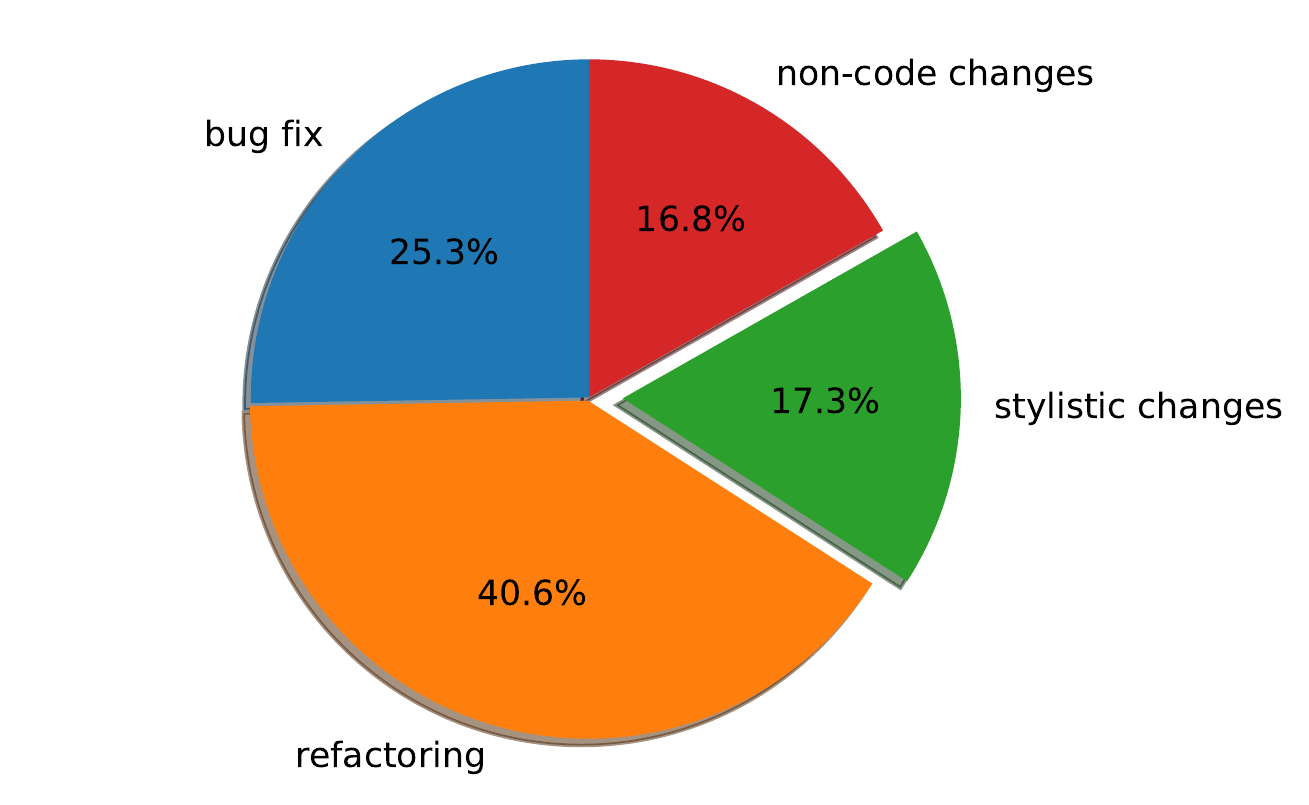}
\caption{Manually created taxonomy for 501 randomly sampled data from test set.}
\label{fig:type-pi-chart}
\end{figure}


\subsection{Which types of changes can our models correctly predict?}
\label{subsec:factor}

We expect our models to suggest fixes for all types of issues reported in code reviews. To inspect the ability to address different types of issues in real scenarios, we conducted a study. We randomly selected
501 samples from our test set. We present the study outcome on the generated fixes of models $model\_{c}$ and $model\_{cc}$.

Two authors performed the manual categorization of different code reviews. To begin with, they jointly labelled 100 samples by discussing each review. The purpose was to develop a shared understanding and to remove individual bias as much as possible. Based on the understanding, they labelled 100 more samples independently. The Cohen kappa \cite{cohenkappa,cohenkappa2} value is 0.64, which indicates substantial agreement between them. Furthermore, the two authors discussed the reviews where disagreements occurred with other authors and converged to a common ground. Next, the remaining 400 samples were labelled equally by the authors independently.

We categorized the possible code changes in four major classes: 1) Bug Fix~\cite{chen2018sequencer, tufano2019empirical, codit} 2) Refactoring~\cite{tufano2019learning} 3) Stylistic change (changes related to indentation and formatting) and 4) Non-code change (changes in documentation and annotation). This is the first work to consider and repair changes that belong to the last two categories to the best of our knowledge. Figure~\ref{fig:type-pi-chart} presents the distribution of the major classes. Note that these two categories cover 34\% changes in our dataset. We show how our models $model\_{c}$ ~\& $model\_{cc}$ perform for changes of different categories in Table~\ref{tab:taxonomy} when we consider top-10 accuracy. We illustrate some of the successfully generated examples by $model\_{cc}$ for each major class sub-categories.



\subsubsection{Bug Fix}

This category consists of code changes that are necessary to overcome system glitch, incorrect output, and unwanted behavior ~\cite{bug_def}. We observe a total of 14 sub-categories under Bug Fix. In the successful case illustrated in Table \ref{tab:bug_1} from the project GoogleReview, we see the reviewer shows an argument why the exception should be thrown conditionally as a high-level overview. Our model successfully generates the target code as specified by the reviewer.

\begin{table}[ht]
\vspace{-1mm}
\centering
\resizebox{\textwidth}{!}{%
\begin{tabular}{ll}
\hline
\multicolumn{2}{l}{\begin{tabular}[c]{@{}l@{}}\textbf{Code Review}: Should this be done conditionally? Otherwise it would try to update username even \\ though it is already set? Or am I missing something?\end{tabular}} \\ \hline
\multicolumn{1}{l|}
{\begin{tabular}[c]{@{}l@{}}
\textbf{Code Before Change:} \\ \texttt{try }\{..\} \\\texttt{catch (OrmDuplicateKeyException dupeErr)}\{  \\~~\textcolor{red}{\texttt{if (!other.isPresent() || }} \\ \textcolor{red}{\texttt{!other.get().accountId().equals(accountId))} \{}  \\~~~~\texttt{throw new IllegalArgumentException("username "} \\ \texttt{ + username + " already in use")}; \\~~\textcolor{red}{\texttt{\}}}  \\ \texttt{\}}
\end{tabular}} &
\begin{tabular}[c]{@{}l@{}}
\textbf{Code After Change:} \\ \texttt{try }\{..\} \\ \texttt{catch (OrmDuplicateKeyException dupeErr)}\{ \\ ~~~~\texttt{throw new IllegalArgumentException("username "} \\ \texttt{+ username + " already in use")};\\~~\texttt{\}} \\
\end{tabular} \\ \hline
\end{tabular}%
}
\vspace{-2mm}
\caption{An example of bug-fix successfully generated by our $model\_cc$. Deleted tokens are marked 
by red color.}
\label{tab:bug_1}
\vspace{-2mm}
\end{table}

\subsubsection{Refactoring}
Refactoring includes code changes intended for code maintenance, which does not change external behavior \cite{refactor_def} of the system. We observe a total of 23 sub-categories under Refactoring. We illustrate a successful sample generated by $model\_{cc}$ from the project Android. 

\begin{table}[ht]
\centering
\resizebox{\textwidth}{!}{%
\begin{tabular}{ll}
\hline
\multicolumn{2}{l}{\begin{tabular}[c]{@{}l@{}}\textbf{Code Review}: This is redundant: If the right hand side of  the '||' operator is evaluated, it is because \texttt{ni != null}. \end{tabular}} \\ \hline
\multicolumn{1}{l|}
{\begin{tabular}[c]{@{}l@{}}
\textbf{Code Before Change:} \\
\texttt{if (ni == null || \textcolor{red}{(ni != null \&\&} !ni.isConnected()\textcolor{red}{)}) \{} \\
~~\texttt{if (LOGD)} \\
~~~~\texttt{~~~~Log.d(TAG, "forceRefresh: no connectivity");} \\
~~\texttt{return false;} \\
\}
\end{tabular}} &
\begin{tabular}[c]{@{}l@{}}
\textbf{Code After Change:} \\ 
\texttt{if (ni == null || !ni.isConnected()) \{} \\
~~\texttt{if (LOGD)}\\
~~~~\texttt{~~~~Log.d(TAG, "forceRefresh: no connectivity");} \\
~~\texttt{return false;} \\
\} \\
\end{tabular} \\ \hline
\end{tabular}%
}
\vspace{-2mm}
\caption{An example of refactoring successfully generated by our $model\_cc$. Deleted tokens are marked 
by red color.}
\label{tab:refac_1}
\vspace{-4mm}
\end{table}


\subsubsection{Stylistic Change}
Code changes required to ensure proper indentation and formatting such as newline insertion, tab spacing, whitespace addition/deletion are considered under this category~\cite{indentation_def}.  We illustrate this with two successful test samples. In the first example, the reviewer emphasizes to add whitespace before \texttt{BluetoothDevice.PHY\_LE\_2M}. Our model generates a correct solution by adding a whitespace token between \texttt{!=} and \texttt{BluetoothDevice.PHY\_LE\_2M}. Similarly, a second example breaks a long line into two lines as the reviewer commented.

\begin{table}[ht]
\vspace{-1mm}
\centering
\resizebox{\textwidth}{!}{%
\begin{tabular}{ll}
\hline
\multicolumn{2}{l}{\begin{tabular}[c]{@{}l@{}}\textbf{Code Review:} Missing space\end{tabular}} \\ \hline
\multicolumn{1}{l|}{\begin{tabular}[c]{@{}l@{}}\textbf{Code Before Change:} \\ 
\texttt{public Builder setSecondaryPhy(int secondaryPhy) \{} \\ 
~~\texttt{if (secondaryPhy != BluetoothDevice.PHY\_LE\_1M \&\&} \\ 
~~\texttt{ ~ secondaryPhy \mybox[fill=red!20]{!=BluetoothDevice.PHY\_LE\_2M} \&\&} \\  
~~\texttt{ secondaryPhy != BluetoothDevice.PHY\_LE\_CODED)}\{...\}\\...\}
\end{tabular}} &
\begin{tabular}[c]{@{}l@{}}
\textbf{Code After Change:} \\ 
\texttt{public Builder setSecondaryPhy(int secondaryPhy) \{} \\ 
~~\texttt{if (secondaryPhy != BluetoothDevice.PHY\_LE\_1M \&\&} \\ 
~~\texttt{ secondaryPhy != BluetoothDevice.PHY\_LE\_2M \&\&} \\  
~~\texttt{ secondaryPhy != BluetoothDevice.PHY\_LE\_CODED)}\{...\}\\...\}\end{tabular} \\ \hline \vspace{-4mm} \\ \hline
\multicolumn{2}{l}{\begin{tabular}[c]{@{}l@{}}\textbf{Code Review:} Long line\end{tabular}} \\ \hline
\multicolumn{1}{l|}{\begin{tabular}[c]{@{}l@{}}\textbf{Code Before Change:} \\ \texttt{public Bundle createTransportModeTransform} \\ \texttt{(IpSecConfig c, IBinder binder) throws RemoteException \{..\}}\end{tabular}} & \begin{tabular}[c]{@{}l@{}}\textbf{Code After Change:} \\ \texttt{public Bundle createTransportModeTransform} \\ \texttt{(IpSecConfig c, IBinder binder)}\mybox[fill=green!20]{~~~~~~~~~~~~~~~~~~~~}\\ \mybox[fill=green!20]{~~~} \texttt{throws RemoteException \{..\}}\end{tabular} \\ \hline
\end{tabular}%
}
\vspace{-2mm}
\caption{Two examples of stylistic change successfully generated by our $model\_cc$. 
The red highlighted portion indicates the code region where a space was added between the tokens and
The green highlighted portion indicates the code region where a newline was added.}
\label{tab:sty_chan}
\vspace{-2mm}
\end{table}

\subsubsection{Non-code change}
We consider changes in non-code regions such as string value, log, code comment, documentation, annotation, and copyright license header under this category~\cite{code_comment,documentation}. We analyze an example of this category, where the reviewer mentions `2017' as the appropriate copyright license year for the file. 
Our model was able to capture the domain-specific context and generate the intended target in this case. Similarly, the second example removes an unnecessary annotation.

\begin{table}[ht]
\centering
\resizebox{\textwidth}{!}{%
\begin{tabular}{ll}
\hline
\multicolumn{2}{l}{\begin{tabular}[c]{@{}l@{}}\textbf{Code Review:} 2017 \end{tabular}} \\ \hline
\multicolumn{1}{l|}{\begin{tabular}[c]{@{}l@{}}
\textbf{Code Before Change:} \\ \texttt{* Copyright (C) \textcolor{red}{2016} The Android Open Source Project}
\end{tabular}} &
\begin{tabular}[c]{@{}l@{}}
\textbf{Code After Change:} \\ \texttt{* Copyright (C) \textcolor{green}{2017} The Android Open Source Project}\end{tabular} \\ \hline \vspace{-4mm} \\ \hline
\multicolumn{2}{l}{\begin{tabular}[c]{@{}l@{}}\textbf{Code Review:} Is this really nullable? What's the point of doing a ref-update validator without a refdb?\end{tabular}} \\ \hline
\multicolumn{1}{l|}{\begin{tabular}[c]{@{}l@{}}\textbf{Code Before Change:} \\ \texttt{public BatchRefUpdateValidator}  \\
      \texttt{(SharedRefDatabase sharedRefDb, 
      ...., } \\ \texttt{\textcolor{red}{@Nullable} @Assisted RefDatabase refDb)} \\ \texttt{\{...\}}\end{tabular}} & \begin{tabular}[c]{@{}l@{}}\textbf{Code After Change:} \\ \texttt{public BatchRefUpdateValidator}  \\
      \texttt{(SharedRefDatabase sharedRefDb, 
      ...., } \\ \texttt{@Assisted RefDatabase refDb)} \\ \texttt{\{...\}} \end{tabular} \\ \hline
\end{tabular}%
}
 \vspace{-2mm}
\caption{Two examples of non-code change successfully generated by our $model\_cc$. Deleted and inserted tokens are marked 
by red and green color respectfully.}
\label{tab:non_code}
\vspace{-4mm}
\end{table}

\section{Ablation Study}
\label{sec:ablation_study}

\begin{table}[ht]
\centering
\hspace*{-2mm}
\resizebox{0.8\textwidth}{!}{%
\begin{tabular}{c|lcccc}
\toprule
  & \textbf{ID} & \textbf{Modified property}   & \textbf{Top-1} & \textbf{Top-5}   & \textbf{Top-10} \\ \midrule
\multirow{7}{*}{\rotatebox[origin=c]{90}{\textit{\textbf{model\_cc}}}}
 & \textbf{1}  & \textbf{Baseline \textit{model\_cc}}   & -   & -   & -   \\ \cline{2-6} 
  & 2  &  Soft tokenization   & -3.03   & \textbf{+5.33}   & \textbf{+11.22} \\ \cline{2-6} 
  & 3  &  Without custom vocabulary selection   & -3.10   & -12.91   & -16.82 \\ \cline{2-6} 
  & 4  &  \begin{tabular}[c]{@{}l@{}}Larger vocabulary \\(10k from code, 10k from CR)\end{tabular}   & -3.44   & -9.74   & -11.89   \\ \cline{2-6} 
  & 5  &  \begin{tabular}[c]{@{}l@{}}Smaller vocabulary \\ (1k from code, 5k from CR)\end{tabular}   & -7.06   & -3.89   & -1.82   \\ \cline{2-6} 
  & 6  &  Smaller embedding size (128)   & -4.65   & -2.8   & -1.82   \\ \cline{2-6} 
  & 7  &  With coverage mechanism   & -2.93   & -4.01   & -2.46   \\ \midrule
\multirow{6}{*}{\rotatebox[origin=c]{90}{\textit{\textbf{model\_c}}}}  
  & \textbf{8} & \textbf{Baseline \textit{model\_c}}   & -   & -   & -   \\ \cline{2-6} 
  & 9  &  Soft tokenization   & -10.81   & -4.2   & -1.87   \\ \cline{2-6} 
  & 10  &  Larger vocabulary (10k from code)   & +0.82   & -3.7   & -4.47   \\ \cline{2-6} 
  & 11  &  \begin{tabular}[c]{@{}l@{}}Smaller vocabulary (1k from code) \\ and smaller embedding size (128)\end{tabular} & -1.65   & -2.25   & -4.04   \\  
\bottomrule
\end{tabular}
}
\caption{Comparison with different network parameters and properties.}
\label{tab:ablation}
\end{table}

We perform an ablation study to understand the relative importance of each design choice for our model. We show these results in Table \ref{tab:ablation}. We define our final models as \textit{baselines} in the table (please see section \ref{subsec:Parameters} for the details about the hyperparameters of these two models). We have two major modifications in our \textit{baselines} including: customized vocabulary (section~\ref{subsubsec:vocab}) and exclusion of coverage mechanism (section~\ref{subsec:network}). To analyze the effect of the coverage mechanism, we train a model with coverage mechanism enabled. The result shows that performance drops after enabling coverage mechanism (ID7). This might be because coverage penalizes repetition of tokens in output, whereas programs usually have repetitions. Without custom vocabulary, the model's performance also decreases (ID3). We consider the review comment and code separately in our custom vocabulary and take the most frequent tokens from them separately. Whereas in ID3, review comment and code vocabularies are merged and most frequent tokens are taken from them jointly. The accuracy decreases in ID3 because the programming language tokens are larger in frequency, so the natural language tokens are not much prevalent in the vocabulary of ID3. 
In Table \ref{tab:ablation}, ID5 and ID11 show that a smaller vocabulary than \textit{baselines} perform worse. ID4 and ID10 show that a larger vocabulary performs poorly as well. Thus, it is clear that our baseline models adopt the best choice of vocabulary size. ID6 and ID11 show that a smaller embedding size than 256 performs worse.




\section{Threats to validity}

In this section, we discuss possible threats that may affect our methodology and the measures taken to mitigate them.

\textbf{Internal validity:} Overlapping data in training and test set is a major problem in deep learning-based source code analysis. After data has been mined from the repositories, we ensured that all <code before the change, code review, code fix> tuples in our dataset are unique. The training and test dataset were created after the deduplication process. Furthermore, the latest 5\% data from each project was selected as the test data, so training and test data are from different time periods. Thus there is no overlap between training and test data.

There is a possibility that the projects using other code review tools (e.g., ReviewBoard, Github pull-based reviews, and Phabricator) might make a difference in our tool's performance. However, we do not use any feature exclusive to Gerrit only, and most of the code review tools' basic workflow is very much similar. Hence, we believe that this threat is minimal.


\textbf{External validity:}
Another threat to the validity of our project is the availability of the code review comments. Also, how will the models operate without the review comments? First of all, our $model\_c$ can operate on bugs without the review comments, and we observed that it achieved similar performance to other approaches on their dataset~\cite{chen2018sequencer,tufano2019learning}. Note that all the learning-based tools evaluated using an oracle (perfect) fault-localizer to have a fair comparison. While applying to a real scenario, our $model\_c$ can also utilize similar fault-localizers used by others. Therefore, not having the review comments does not make our tool inapplicable at all. Secondly, Table~\ref{tab:project-distribution} shows that we have around 950K code reviews from 15 projects. Code review comments are prevalent, and all the code reviews need to be addressed before merging to the main codebase or abandoning the code. If we can reduce the developers' effort for a significant amount of the review requests, that will reduce software developers' burden.

 

\label{subsec:threats}

\section{Related Work}
\label{sec:related-works}
Automatic Program Repair (APR) is an active area of research for ages. In recent years, large companies such as Facebook have started such tools in the production environment \cite{bader2019getafix}. Along with classic approaches of dynamic and static analysis based repair, Machine Learning based techniques are also showing immense promise.  

\textbf{Classic Automatic Program Repair (APR):}
Researchers have been trying to automatically repair software systems by generating an actual fix for more than two decades \cite{survey}. Genprog \cite{genprog-1, genprog-2, genprog-3} is a Genetic Algorithm based automated repair technique using test suites. Arcuri \cite{Arcuri}, Debroy and Wong \cite{debroy2010using}, Kern and Esparza \cite{kern2010automatic} proposed mutation-based repair techniques. SemFix \cite{semfix} and Angelix \cite{angelix} attempted automated program repair based on symbolic execution. The PAR system \cite{par} automatically fixed java code with ten repair templates. Könighofer and Bloem \cite{konighofer2011automated} considered assertions in programs and used an SMT solver to fill holes in repair templates. Samimi et al. \cite{samimi2012automated} repaired PHP programs using correctly generated HTML format. Liu et al. \cite{liu} parameterized a manually written bug report and extracted necessary values from the report to repair the program. Many program repair methods used a reference implementation for repair~\cite{konighofer2012repair, singh2013automated}. Many methods have been tried in the literature for program repair, whether as test suites or formal restrictions. However, before the recent advancement in Deep Learning and Natural Language Processing, using an unstructured natural language text for program repair was an unthinkable concept.

\textbf{Machine Learning-based Automatic Program Repair:} Applying a Deep learning-based approach to detect and fix bugs has shown promising results in recent years, mainly due to the availability of large datasets. Pu \textit{et al.} ~\cite{skip} proposed a seq2seq network for automatic program correction in MOOCs. Hata \textit{et al.} \cite{hata2018learning} performed automatic patch generation with neural machine translation. ENCORE \cite{encore} used an ensemble of multiple Convolutional Neural Network \cite{cnn} based Neural Machine Translation models to improve the performance of Deep Learning-based APR.

In two consequent papers, Tufano \textit{et al.}~\cite{tufano2019empirical, tufano2019learning} empirically demonstrated the applicability of NMT for program repair. They released a test dataset named Bugs2Fix and achieved 9\% accuracy in it \cite{tufano2019empirical}. They also applied the seq2seq model with attention mechanism \cite{bahdanau} for repairing Java functions that are less than 50 tokens or 50 to 100 token long after applying their proposed \cite{tufano2019learning} tokenization method.

SequenceR \cite{chen2018sequencer} proposed an abstraction method to capture contexts from the source code. Their model can fix in-line bugs inside functions with 20\% accuracy on Bugs2Fix dataset. They also demonstrated how Copy mechanism~\cite{chen2018sequencer} can be used to solve infinite vocabulary problem~\cite{infinitevar} for program repair. Our study has a much broader context than both Tufano \textit{et al.} \cite{tufano2019empirical, tufano2019learning} and SequenceR \cite{chen2018sequencer} since we can perform multiline changes inside/outside functions and can handle functions of any size.

Getafix \cite{bader2019getafix} is a tool developed and internally used by Facebook. It firstly splits a given set of recurring example fixes into AST-level edits. By applying the agglomerative clustering technique, this algorithm produces a hierarchy of fix patterns where the child nodes produce the most specific fixes. The higher the nodes are in the hierarchy; the patterns get more and more generalized. Finally, given a bug under fix, it finds the most suitable fix patterns from that hierarchy, ranks candidate fixes, and suggests developers' top-most fixes.

CODIT \cite{codit} models code changes with tree-based machine translation. Instead of the source code, they work on the underlying syntax tree of the code. They divide the task of predicting code changes in two steps: Firstly, they learn and predict the edited code's syntax tree. Secondly, given the predicted tree structure, they generate the tokens, i.e., variables, keywords etc. using the Seq2seq mechanism.

HOPPITY \cite{hoppity} models the problem of bug-fixing as learning a sequence of graph transformations. Given a buggy program modelled by a graph structure, their approach makes a sequence of predictions to identify bug nodes' position and corresponding graph edits to produce a fix. To model the graph structure of the source code, they pass the processed syntax tree(AST) through a Graph Neural Network(GNN)~\cite{graph} and produce a fixed dimensional vector space.

DLFix~\cite{dlfix} adopts a two-tier Deep learning model for APR using Tree-LSTM~\cite{tree_LSTM}. First, the changed sub-tree in the AST is summarized in a single node to learn the local context. Second, the summarized node information and the sub-tree difference before and after the change are used to learn the code transformation. Additionally, they also deploy a classification model to re-rank the generated patches.

Although there has been plenty of noticeable works in Machine Learning-based program repair, to the best of our knowledge, none of the previous works has exploited the additional information of code review to improve the performance of the techniques.  Since there is a vast scope of improvement for the quality of suggested fixes, these and the future approaches can be improved using our proposed mechanism of utilizing code review.

\vspace{-4mm}
\section{Conclusions}
\label{sec:conclusions}
This research has increased the quality of fix suggestions by exploiting the code review comments. We have prepared a dataset of 55,060 triplets (code before the change, code after the change, and code review) and show that our approach improves up to 34.82\% over state-of-the-art APR techniques. This is the first step towards learning to repair programs using Natural Language instructions to the best of our knowledge. We systematically analyzed the generated fix suggestions and found some categories left out by other techniques.

In the future, we will aim to increase our model's performance by harnessing advanced Deep Learning architecture, such as using different encoders for code review and code or using pre-trained encoders such as BERT \cite{bert}.  We hope that the released dataset and code will help research in Automatic Program Repair using code review.

\section{Acknowledgement}
The research was partially supported by a grant "Code Review Measurement" granted by Samsung Research Bangladesh.

\bibliography{main}

\newpage

\begin{appendix}
\section{Taxonomy for top-10 predictions}
\label{tab:taxonomy}
\begin{figure}[ht]
\centering
\scalebox{0.50}{
\includegraphics[width=.88\linewidth]{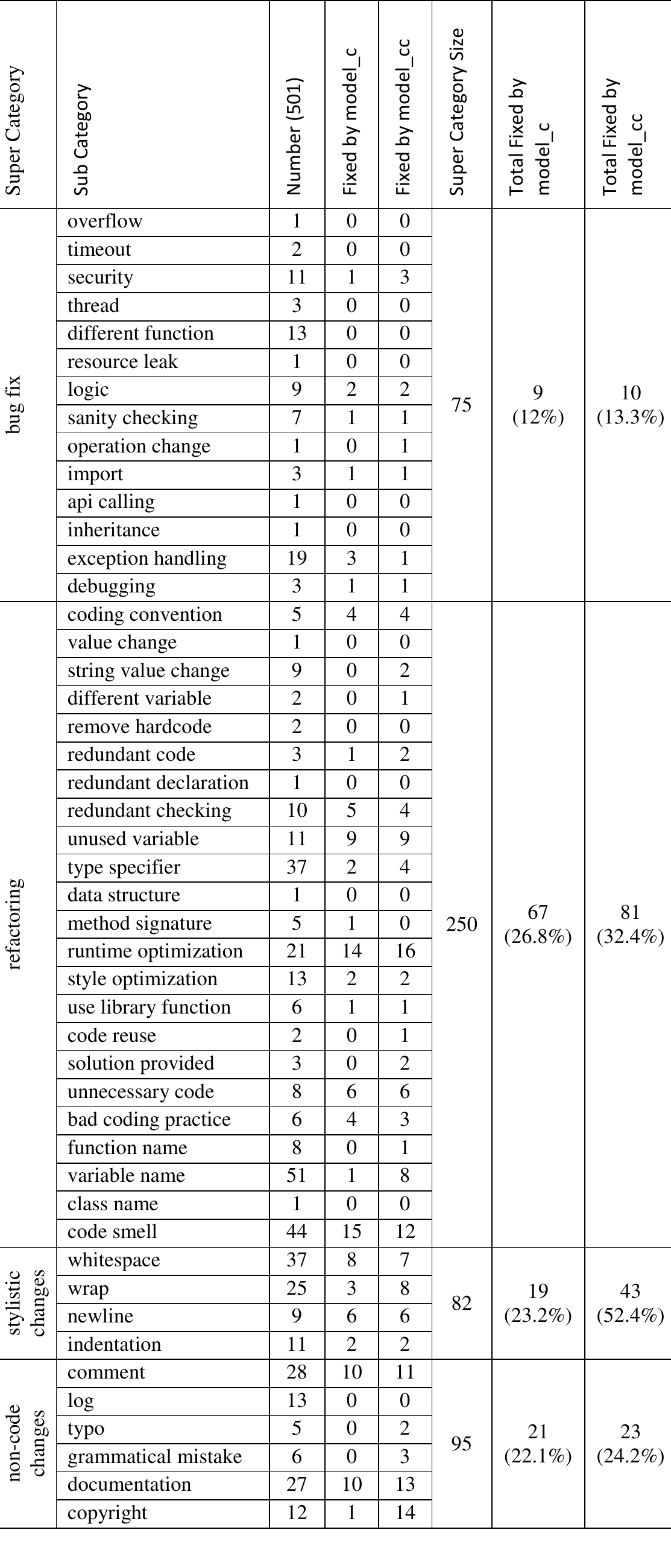}
}
\end{figure}

\end{appendix}

\end{document}